\documentclass[12pt]{article}
\pdfoutput=1

\usepackage{draft} 
\usepackage{hyperref}
\usepackage{graphicx,color,subfig}
\usepackage{cite}
\usepackage{mciteplus}
\usepackage{skak}
\usepackage{empheq}
\usepackage{tikz}
\usepackage{booktabs}
\usepackage{siunitx}
\usepackage{bbm}
\usepackage{makecell}
\usepackage{float}
\usepackage{dsfont}

\usepackage{bm}
\usepackage{braket}

\DeclareFontFamily{OT1}{pzc}{}
\DeclareFontShape{OT1}{pzc}{m}{it}{<-> s * [1.10] pzcmi7t}{}
\DeclareMathAlphabet{\mathpzc}{OT1}{pzc}{m}{it}

\def\be#1\ee{\begin{align}#1\end{align}}

\makeatletter
\newcommand{\bdryno}{\mathpalette\bdry@no\relax}
\newcommand{\bdry@no}[2]{%
  \mspace{1mu}%
  \vbox{%
    \hbox{$\m@th#1\scriptstyle{\ast}$}
    \nointerlineskip
    \kern.25ex
    \hbox{$\m@th#1\scriptstyle{\ast}$}
    \kern-.06ex
  }%
  \mspace{1mu}%
}
\makeatother

\usetikzlibrary{snakes}
\usetikzlibrary{arrows}
\usetikzlibrary{decorations.pathmorphing}
\usetikzlibrary{ decorations.markings}
\tikzset{snake it/.style={decorate, decoration=snake}}
\usetikzlibrary{shapes.misc}
\tikzset{cross/.style={cross out, draw=black, minimum size=2*(#1-\pgflinewidth), inner sep=0pt, outer sep=0pt},
cross/.default={1pt}}
\usetikzlibrary{shapes.geometric}

\newcommand{\nssect}{\text{NS}}
\newcommand{\rsect}{\text{R}}
\newcommand{\cdtwo}{C_{D^2}}

\newcommand{\Dbar}{\overline{D}}

\newcommand{\cT}{\mathcal{T}}
\newcommand{\cR}{\mathcal{R}}

\DeclareMathOperator*{\SumInt}{%
\mathchoice%
  {\ooalign{$\displaystyle\sum$\cr\hidewidth$\displaystyle\int$\hidewidth\cr}}
  {\ooalign{\raisebox{.14\height}{\scalebox{.7}{$\textstyle\sum$}}\cr\hidewidth$\textstyle\int$\hidewidth\cr}}
  {\ooalign{\raisebox{.2\height}{\scalebox{.6}{$\scriptstyle\sum$}}\cr$\scriptstyle\int$\cr}}
  {\ooalign{\raisebox{.2\height}{\scalebox{.6}{$\scriptstyle\sum$}}\cr$\scriptstyle\int$\cr}}
}

\tikzset{
  pics/cylnnrb0/.style n args={2}{
    code = { %
        \filldraw[color=red,fill=black!30, thick] circle (0.35);
        \filldraw[color=blue, fill=white, thick]  circle (0.17);
        \node[color=black] at (0,0.51) {\scriptsize ${#1}$};
        \node[color=black] at (0,0) {\scriptsize ${#2}$};

    }
  }
}

\tikzset{
  pics/cylnnrb/.style n args={2}{
    code = { %
        \filldraw[color=red, densely dashed,fill=black!30, thick] circle (0.35);
        \filldraw[color=blue, densely dashed, fill=white, thick]  circle (0.17);
        \node[color=black] at (0,0.51) {\scriptsize ${#1}$};
        \node[color=black] at (0,0) {\scriptsize ${#2}$};

    }
  }
}
\tikzset{
  pics/cylnnrr0/.style n args={2}{
    code = { %
        \filldraw[color=red,fill=black!30, thick] circle (0.35);
        \filldraw[color=red, fill=white, thick]  circle (0.17);
        \node[color=black] at (0,0.51) {\scriptsize ${#1}$};
        \node[color=black] at (0,0) {\scriptsize ${#2}$};

    }
  }
}
\tikzset{
  pics/cylnnrr/.style n args={2}{
    code = { %
        \filldraw[color=red, densely dashed,fill=black!30, thick] circle (0.35);
        \filldraw[color=red, densely dashed, fill=white, thick]  circle (0.17);
        \node[color=black] at (0,0.51) {\scriptsize ${#1}$};
        \node[color=black] at (0,0) {\scriptsize ${#2}$};

    }
  }
}
\tikzset{
  pics/cylnnrr1/.style n args={2}{
    code = { %
        \filldraw[color=red,fill=black!30, thick] circle (0.35);
        \filldraw[color=red, fill=white, thick]  circle (0.17);
        \node[color=black] at (0,0.51) {\scriptsize ${#1}$};
        \node[color=black] at (0,0) {\scriptsize ${#2}$};
        \node[cross=3pt, very thick] at (-0.25,0) {};
    }
  }
}
\tikzset{
  pics/cylnnbb0/.style n args={2}{
    code = { %
        \filldraw[color=blue,fill=black!30, thick] circle (0.35);
        \filldraw[color=blue, fill=white, thick]  circle (0.17);
        \node[color=black] at (0,0.51) {\scriptsize ${#1}$};
        \node[color=black] at (0,0) {\scriptsize ${#2}$};

    }
  }
}
\tikzset{
  pics/cylnnbb/.style n args={2}{
    code = { %
        \filldraw[color=blue, densely dashed,fill=black!30, thick] circle (0.35);
        \filldraw[color=blue, densely dashed, fill=white, thick]  circle (0.17);
        \node[color=black] at (0,0.51) {\scriptsize ${#1}$};
        \node[color=black] at (0,0) {\scriptsize ${#2}$};

    }
  }
}
\tikzset{
  pics/cylnnbb1/.style n args={2}{
    code = { %
        \filldraw[color=blue,fill=black!30, thick] circle (0.35);
        \filldraw[color=blue, fill=white, thick]  circle (0.17);
        \node[color=black] at (0,0.51) {\scriptsize ${#1}$};
        \node[color=black] at (0,0) {\scriptsize ${#2}$};
        \node[cross=3pt, very thick] at (-0.25,0) {};
    }
  }
}

\tikzset{
  pics/disk1/.style n args={1}{
    code = { %
        \filldraw[color=black, fill=black!30, thick] circle (0.3);
        \node[color=black] at (0,0.5) {\scriptsize ${#1}$};
        \draw node[cross=3pt, very thick] {};
    }
  }
}   
\tikzset{
  pics/disk1r/.style n args={1}{
    code = { %
        \filldraw[color=red, fill=black!30, thick] circle (0.3);
        \node[color=black] at (0,0.5) {\scriptsize ${#1}$};
        \draw node[cross=3pt, very thick] {};
    }
  }
}
\tikzset{
  pics/disk2r/.style n args={1}{
    code = { %
        \filldraw[color=red, fill=black!30, thick] circle (0.3);
        \node[color=black] at (0,0.5) {\scriptsize ${#1}$};
        \node[cross=3pt, very thick] at (0.125,0) {};
        \node[cross=3pt, very thick] at (-0.125,0) {};
    }
  }
}
\tikzset{
  pics/disker/.style n args={1}{
    code = { %
        \filldraw[color=red, fill=black!30, thick] circle (0.3);
        \node[color=black] at (0,0.5) {\scriptsize ${#1}$};
    }
  }
}
\tikzset{
  pics/disk1b/.style n args={1}{
    code = { %
        \filldraw[color=blue, fill=black!30, thick] circle (0.3);
        \node[color=black] at (0,0.5) {\scriptsize ${#1}$};
        \draw node[cross=3pt, very thick] {};
    }
  }
}   
\tikzset{
  pics/disk2b/.style n args={1}{
    code = { %
        \filldraw[color=blue, fill=black!30, thick] circle (0.3);
        \node[color=black] at (0,0.5) {\scriptsize ${#1}$};
        \node[cross=3pt, very thick] at (0.125,0) {};
        \node[cross=3pt, very thick] at (-0.125,0) {};
    }
  }
} 
\tikzset{
  pics/diskeb/.style n args={1}{
    code = { %
        \filldraw[color=blue, fill=black!30, thick] circle (0.3);
        \node[color=black] at (0,0.5) {\scriptsize ${#1}$};
    }
  }
}

\begin{document}

\unitlength = .8mm

\begin{titlepage}

\begin{center}

\hfill \\
\hfill \\
\vskip 1cm

\title{The S-Matrix of 2D Type 0B String Theory
\\
Part 2: D-Instanton Effects}

\author{Bruno Balthazar$^\diamondsuit$, Victor A. Rodriguez$^{\heartsuit}$, Xi Yin$^\spadesuit$}

\address{
$^\diamondsuit$Enrico Fermi Institute \& Kadanoff Center for Theoretical Physics,\\
University of Chicago, Chicago, IL 60637, USA \\
$^\heartsuit$Joseph Henry Laboratories, Princeton University, \\ Princeton, NJ 08544, USA \\
$^\spadesuit$Jefferson Physical Laboratory, Harvard University, \\
Cambridge, MA 02138 USA
}

\email{brunobalthazar@uchicago.edu, vrodriguez@princeton.edu, xiyin@fas.harvard.edu}

\end{center}

\abstract{We study the effect of D-instantons on closed string scattering amplitudes in the two-dimensional type 0B string theory from the worldsheet perspective. We find that the contribution from a pair of D-instanton and anti-D-instanton to the closed string reflection amplitude, with a suitable contour prescription for the integration over the D-instanton moduli space, agrees with the corresponding leading non-perturbative corrections in the proposed dual matrix quantum mechanics. This analysis is further extended to thermal observables defined at finite temperature. The infrared divergence in charged processes is understood through the measure factor for charged D-instantons, and can be treated with spacetime dimensional regularization. }

\vfill

\end{titlepage}

\eject

\begingroup
\hypersetup{linkcolor=black}

\tableofcontents

\endgroup

\section{Introduction}


The duality between the two-dimensional type 0B string theory and a scaling limit of a gauged matrix quantum mechanics (MQM) has been conjectured in \cite{Takayanagi:2003sm, Douglas:2003up} and explored at the level of string perturbation theory in \cite{Balthazar:2022atu}. In this paper we study the non-perturbative corrections to the closed string S-matrix in this setting, in particular the effect of D-instantons.

The worldsheet theory of type 0B string in the Neveu-Schwarz-Ramond formalism is a superconformal field theory that involves ${\cal N}=1$ Liouville theory and a diagonal GSO projection. The perturbative string excitations consist of two types of particles propagating in the two-dimensional target spacetime, both massless scalars in the asymptotic region, one from the (NS, NS) sector known as the ``tachyon", the other from the (R, R) sector known as the ``axion". The asymptotic string states are dual to, in the matrix model description, collective excitations of the fermi surface of free non-relativistic fermions subject to the Hamiltonian $h={1\over 2}(p^2-x^2)$. At the perturbative or semi-classical level, the fermi sea filling the phase space up to energy $-\mu$ ($\mu>0$) consists of two disconnected components, one at $x>0$ (``right side") and the other at $x<0$ (``left side"). 

Non-perturbatively, the collective excitations of the left and right sides of the fermi surface do interact with one another due to the tunneling of fermions through the potential barrier. 
The full Hilbert space of type 0B string theory is expected to be spanned by asymptotic closed string states in ``charged sectors", of the form
\ie
{\cal H} \simeq \bigoplus_{k\in\mathbb{Z}} {\cal H}_k^{\rm in}  \simeq \bigoplus_{k\in\mathbb{Z}} {\cal H}_k^{\rm out} ,
\fe
where the charge $k$ may be interpreted as the difference between the numbers of fermions in the left versus right asymptotic regions of the fermi sea. As the total number of fermions is infinite, all ${\cal H}_k$'s are isomorphic to one another. On the other hand, a tunneling event of a fermion from the RHS to the LHS (or vice versa) changes $k$ by $+1$ (or $-1$). The full S-matrix is expected to be graded according to
\ie
{\cal S} = \bigoplus_{\ell\in\mathbb{Z}} {\cal S}_\ell,
\fe
where ${\cal S}_\ell$ maps ${\cal H}_k$ to ${\cal H}_{k+\ell}$. The perturbative closed string S-matrix amounts to the perturbative expansion of the ``uncharged sector" S-matrix ${\cal S}_0$ in $g_s$, or equivalently in $1/\mu$. ${\cal S}_0$ is well-defined non-perturbatively and admits a D-instanton expansion as we will review in section \ref{sec:unchargedmqmamp}. Note that ${\cal S}_0$ by itself does not saturate unitarity.

The ``charged sector" S-matrix ${\cal S}_{\ell\not=0}$, on the other hand, suffers from infrared divergence as was pointed out in \cite{DeWolfe:2003qf}. This is quite reminiscent of the scattering amplitudes involving charged particles in a non-confining gauge theory in four dimensions, where the individual S-matrix elements vanish, while the inclusive scattering probabilities taking into account soft radiation are finite. In the type 0B MQM, the IR divergence in the charged sector indicates that the Fock space basis of asymptotic collective modes of the fermi surface is not suitable for describing processes that involve a nonzero net number of tunneled fermions.

\begin{figure}[h!]
\centering
\begin{tabular}{c c c}
\scalebox{0.75}{
\begin{tikzpicture}
\fill[black!30] (-5,-4) rectangle (5,-3/2);
\draw[color=black, fill=black!30, line width=2pt] (1.225,-3/2) -- (5,-3/2);
\draw[color=black, fill=black!30, line width=2pt] (-1.225,-3/2) -- (-5,-3/2);
\draw[black, line width=2pt, fill=white, domain=-2.005:2.005] plot (\x, {-\x*\x});
\draw[->,color=red, line width=2pt] (1.1,-1.5) -- (-1.1,-1.5);
\node[color=red] at (0, -1.9) {D};
\end{tikzpicture}}
&&
\scalebox{0.75}{\begin{tikzpicture}
\fill[black!30] (-5,-4) rectangle (5,-3/2);
\draw[color=black, fill=black!30, line width=2pt] (1.225,-3/2) -- (5,-3/2);
\draw[color=black, fill=black!30, line width=2pt] (-1.225,-3/2) -- (-5,-3/2);
\draw[black, line width=2pt, fill=white, domain=-2.005:2.005] plot (\x, {-\x*\x});
\draw[->,color=blue, line width=2pt] (-1.1,-1.5) -- (1.1,-1.5);
\node[color=blue] at (0, -1.9) {$\overline{\rm D}$};
\end{tikzpicture}} \\
D-instanton && $\overline{\rm D}$-instanton 
\end{tabular}
\caption{A D-instanton mediated process corresponds to the tunneling of a particle across the inverted quadratic potential, say from right to left. A $\overline{\rm D}$-instanton process corresponds to the tunneling of a particle across the potential in the opposite direction, or equivalently, to the tunneling of a hole across the potential in the same direction.}
\label{fig:DinstTunneling}
\end{figure}
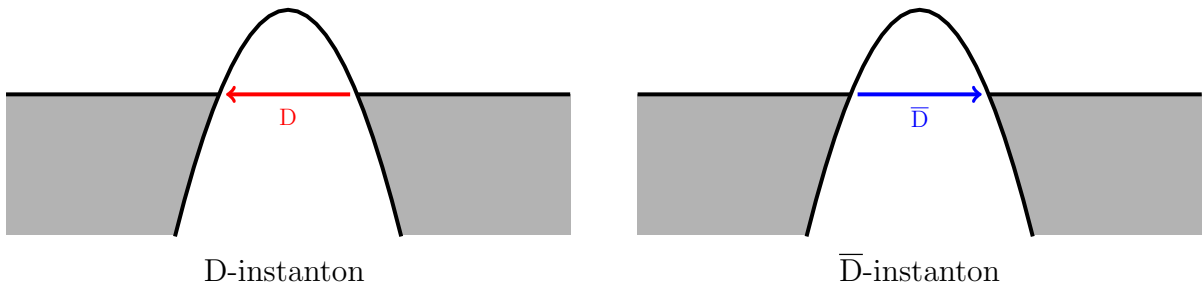

From the worldsheet perspective, non-perturbative effects in type 0B string theory are mediated by D-instantons. The simplest D-instantons can be described as a boundary condition of the worldsheet CFT that is the tensor product of the Dirichlet boundary condition in $(X^0,\psi^0,\widetilde\psi^0)$ and a supersymmetric analog of the $(1,1)$ ZZ boundary condition in the ${\cal N}=1$ Liouville theory (reviewed in section \ref{Dinst}). They carry $\pm1$ unit of RR charge, have action $\pi \mu$, and can be identified with the tunneling of a fermion in the MQM, from the RHS to the LHS of the fermi sea, or vice versa. We will refer to them as the (elementary) D-instanton and $\overline{\rm D}$-instanton (i.e. anti-D-instanton) respectively. 

\begin{figure}[h!]
\centering
\begin{tikzpicture}
\fill[left color = black!0,right color = black!30] (2,-1) -- (7.2,-1) -- (7.2,5.75) -- (2,5.75) -- (2,-1);
\draw[color=white,line width=1pt] (2,-1) -- (7.2,-1) -- (7.2,5.75) -- (2,5.75) -- (2,-1);
\draw[->,color=black, line width=1pt] (-1,0) -- (5,0);
\node[right] at (5,0) {$\phi~$ (space)};
\draw[->,color=black, line width=1pt] (0,-1) -- (0,5);
\node[above] at (0,5) {$X^{0}~$ (time)};
\draw[color=black, line width=1pt] (-0.1,2) -- (0.1,2);
\draw[color=black, line width=1pt] (-0.1,4) -- (0.1,4);
\node[left] at (-0.1,2) {$x^0_1$};
\node[left] at (-0.1,4) {$x^0_2$};
\filldraw[color=gray!40, line width=1pt] (0.5,-1) to[out=45,in=-90] (4,2) to[out=-120,in=45] (0.5,-0.2);
\draw[color=black, line width=1pt] (0.5,-1) to[out=45,in=-90] (4,2) to[out=-120,in=45] (0.5,-0.2);
\draw[color=black, line width=1pt] (1,-0.6) to[out=45,in=-45] (1,0.2);
\draw[color=black,dashed, line width=1pt] (1,-0.6) to[out=130,in=-130] (1,0.2);
\filldraw[color=gray!40, line width=1pt] (4,2) to[out=20,in=-5] (4,4) to[out=-70,in=55] (4,2);
\draw[color=black, line width=1pt] (4,2) to[out=20,in=-5] (4,4) to[out=-70,in=55] (4,2);
\draw[color=black, line width=1pt] (4.23,3.2) to[out=-45,in=-135] (4.57,3.2);
\draw[color=black, dashed, line width=1pt] (4.23,3.2) to[out=45,in=135] (4.57,3.2);
\filldraw[color=gray!40, line width=1pt] (4,2) to[out=160,in=-160] (4,4) to[out=-130,in=110] (4,2);
\draw[color=black, line width=1pt] (4,2) to[out=160,in=-160] (4,4) to[out=-130,in=110] (4,2);
\draw[color=black, line width=1pt] (3.7,2.9) to[out=-45,in=-135] (3.475,2.9);
\draw[color=black, dashed, line width=1pt] (3.7,2.9) to[out=45,in=135] (3.475,2.9);
\filldraw[color=gray!40, line width=1pt] (0.5,5.8) to[out=-15,in=140] (4,4) to[out=110,in=-30] (0.48,7.2) to[out=-90,in=90] (0.48,6.8) to[out=-20,in=140] (3.2,5) to[out=155,in=-30] (0.5,6.4);
\draw[color=black, line width=1pt] (0.5,5.8) to[out=-15,in=140] (4,4) to[out=110,in=-30] (0.48,7.2) to[out=-90,in=90] (0.48,6.8) to[out=-20,in=140] (3.2,5) to[out=155,in=-30] (0.5,6.4);
\filldraw[color=white] (0.5,7.3) -- (0.5,6.7) -- (0.3,6.7) -- (0.3,7.3);
\draw[color=black, line width=1pt] (1,5.65) to[out=45,in=-45] (1,6.1);
\draw[color=black, dashed, line width=1pt] (1,5.65) to[out=130,in=-130] (1,6.1);
\draw[color=black, line width=1pt] (1,6.6) to[out=45,in=-45] (1,6.9);
\draw[color=black, dashed, line width=1pt] (1,6.6) to[out=130,in=-130] (1,6.9);
\filldraw[color=gray!40, line width=1pt] (4,4) to[out=100,in=135] (4.7,4.7) to[out=-45,in=-10] (4,4) to[out=0,in=-45] (4.45,4.45) to[out=135,in=90] (4,4);
\draw[color=black, line width=1pt] (4,4) to[out=100,in=135] (4.7,4.7) to[out=-45,in=-10] (4,4) to[out=0,in=-45] (4.45,4.45) to[out=135,in=90] (4,4);
\draw[color=black, line width=1pt] (4.45,4.45) to[out=-20,in=-90] (4.7,4.7);
\draw[color=black, dashed, line width=1pt] (4.45,4.45) to[out=80,in=-170] (4.7,4.7);
\node[cross=6pt, line width=1.5pt] at (4,2) {};
\node[cross=6pt, line width=1.5pt] at (4,4) {};
\end{tikzpicture}
\caption{An artist's impression of a worldsheet configuration in the presence of a pair of D-instantons that contributes to the $1\to 2$ scattering amplitude of closed strings. The region shaded with a gradient represents the Liouville potential wall. The D-instantons that occur at times $x^0_1$ and $x^0_2$ are marked with crosses. }
\label{fig:DinstSpacetime}
\end{figure}
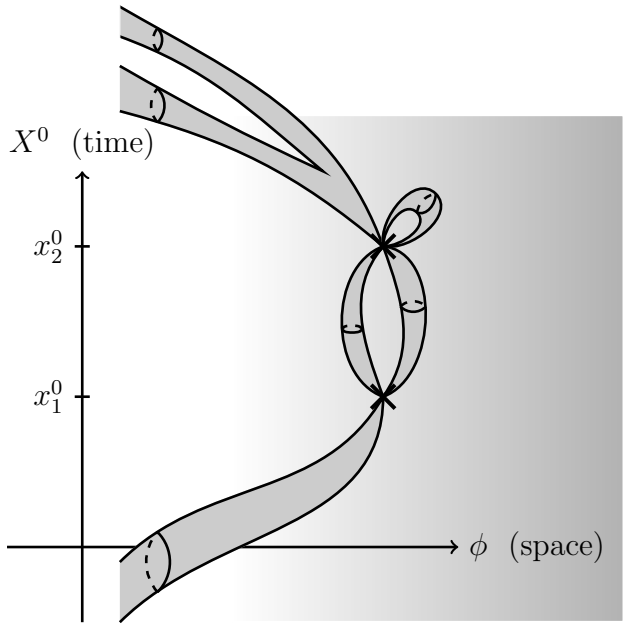

The leading non-perturbative contribution to the uncharged S-matrix ${\cal S}_0$, which occurs at order $e^{-2\pi\mu}$, comes from a pair of D- and $\overline{\rm D}$-instanton. In section \ref{sec:dinstuncharged}, we will explicitly compute the effect of the latter on the $1\to k$ scattering amplitudes of closed string modes, namely those of the tachyon and axion. With a principal value contour prescription in the integration over the instanton moduli space, we obtain results in highly nontrivial agreement with the dual MQM, including at finite temperature.

A single D- or $\overline{\rm D}$-instanton, on the other hand, is expected to contribute to the charged sector S-matrix ${\cal S}_{\pm 1}$ at order $e^{-\pi\mu}$. The latter, as already mentioned, is IR divergent. In the worldsheet description, the origin of this infrared divergence lies in the cylinder diagram that contributes to the measure on the D-instanton moduli space. The divergence can be regularized by working at finite temperature, where the timelike free boson in the worldsheet theory is replaced by a compact boson. We will find that in the zero temperature limit, the measure factor for the charged D- or $\overline{\rm D}$-instanton vanishes, while that of the uncharged D-$\overline{\rm D}$ instanton remains finite.

In this paper, we will also study a generalization of the S-matrix to finite temperature, defined as
\ie\label{finitetsdef}
{\cal S}^{(T)}(\B|\A) =  {\rm Tr}\left[ \rho^{(T)} \prod_{j\in \B} a^{out}_j \prod_{i\in \A} (a_i^{in})^\dagger \right],
\fe
where $\rho^{(T)}$ is the thermal density matrix at temperature $T$, and $a_i^{in}, a_j^{out}$ are annihilation operators of in- and out-particles. 
Usually in quantum field theory, the notion of asymptotic particles is ill-defined at finite temperature. However, in the MQM or 2D non-critical string theories, the effective coupling of the collective modes or closed strings goes to zero in the asymptotic region, which allows for well-defined asymptotic particles in a thermal bath. A notion of thermal amplitudes in this context has been studied previously in perturbation theory in \cite{Klebanov:1991qa}. We will elaborate on the interpretation of such thermal amplitudes and their connection to (\ref{finitetsdef}) via a ``thermal LSZ" relation.

In section \ref{sec:DimReg}, we further consider a dimensional regularization scheme in the target spacetime by deforming the worldsheet CFT. This allows us to access infrared safe observables, namely inclusive scattering probabilities, in the charged processes of type 0B string theory, through dimensionally regularized amplitudes that are manifestly unitary at the D-instanton level. 

We conclude with some future perspectives in section \ref{sec:discuss}.

\section{Results from the matrix quantum mechanics}
\label{sec:mqmreview}

The exact S-matrix of the collective excitations of the type 0B MQM, in the uncharged sector, was solved in \cite{Moore:1991zv}. In this section we recall the derivation of the relevant amplitudes based on LSZ reduction of Green functions for particle/hole pairs, and give formulae for their explicit instanton expansions. We then extend these results to finite temperature.

\subsection{The S-matrix of collective modes from LSZ reduction}
\label{sec:lszred}

We denote by $|n\rangle$ be a basis of single fermion states with energy eigenvalue $E_n$ and wave function $\varphi_n(x)$. For simplicity of exposition we will treat these states as discrete for now, and later take the appropriate continuum limit when necessary. It is convenient to work with second quantized fermion fields
\ie
& \psi(x,t) = \sum_n b_n \varphi_n(x) e^{-iE_nt},
\\
& \psi^\dagger(x,t) =\sum_n b_n^\dagger \varphi_n^*(x) e^{iE_n t},
\fe
where $b_n$ and $b_n^\dagger$ are the annihilation and creation operators for a fermion in the state $|n\rangle$.
Let $|\Omega\rangle$ be the fermi sea ground state in which all energy levels up to $E=-\mu$ are filled. The time-ordered 2-point Green function can be written as
\ie\label{fermiprop}
\langle\Omega| {\bf T} \psi(x_1,t_1) \psi^\dagger(x_2, t_2)  |\Omega\rangle
&= \sum_n e^{-i E_b t_{12}} \varphi_n(x_1) \varphi_n^*(x_2) \Big[ \theta(t_{12}) \theta(E_n+\mu) - \theta(t_{21}) \theta(-E_n-\mu) \Big]
\\
&= {i\over 2\pi} \sum_n \varphi_n(x_1) \varphi_n^*(x_2) \int_{-\infty}^\infty dz\, {e^{-iz t_{12}}\over z - E_n + i\epsilon(E_n+\mu)}.
\fe
The S-matrix of the collective modes is obtained by taking the LSZ limit of the time-ordered Green function of the bosonic field operator $\psi^\dagger\psi(x,t)$. Such Green functions are easily calculated by Wick contractions using the fermion propagator (\ref{fermiprop}). The LSZ limit for a particle/hole pair amounts to extracting the in- or out- components of the asymptotic wave functions $\varphi_n(x) \varphi_m^*(x)$ appearing in the mode expansion of the boson operator $\psi^\dagger\psi(x,t)$.

Take for the instance the $1\to 1$ reflection S-matrix element of a collective mode on the RHS of the fermi sea,
\ie
S_{1_R\to 1_R}(\omega; \omega') = {}^{out}\langle\omega'|\omega\rangle{}^{in} = \delta(\omega-\omega') {\cal A}_{1_R\to 1_R}(\omega).
\label{eq:Smatelem1to1}
\fe
The LSZ reduction of the 2-point time-ordered Green function of $\psi^\dagger\psi$ gives
\ie\label{lszaoo}
{\cal A}_{1_R\to 1_R} (\omega) &= \sum_{n,m} { \left[ \varphi_n \right]_R^{\rightarrow} \left[\varphi_m^*\right]_R^{\rightarrow} } { \left[ \varphi_n^* \right]_R^{\leftarrow} \left[\varphi_m\right]_R^{\leftarrow} }
\int_{-\infty}^\infty dz\, {1\over z -E_n+i\epsilon(E_n+\mu)} {1\over z-\omega - E_m + i\epsilon(E_m-\mu)},
\fe
where $[\varphi_n]_R^\leftarrow$ and $[\varphi_n]_R^\rightarrow$ stand for appropriately normalized coefficients of the incoming and outgoing components of the wave function $\varphi_n(x)$ in the right (hence subscript $R$) asymptotic region.

There is a 2-fold degeneracy at every energy level $E$, corresponding to the fermion scattering wave function $\varphi_E^{(L)}$ with incoming flux from the left, and $\varphi_E^{(R)}$ with incoming flux from the right. The out-components of $\varphi_E^{(R)}$, for instance, are related by
\ie{}
[\varphi_E^{(R)}]_R^\rightarrow = R_p(E) [\varphi_E^{(R)}]_R^\leftarrow,~~~~~ [\varphi_E^{(R)}]_L^\leftarrow = T_p(E) [\varphi_E^{(R)}]_R^\leftarrow,
\fe
where $R_p(E)$ and $T_p(E)$ are the single fermion (``particle") reflection and transmission coefficients, given explicitly by
\ie\label{eq:exactRandT}
R_p(E) = i \mu^{iE} \left[ {1\over 1+e^{2\pi E}}\cdot {\Gamma({1\over 2}-i E)\over \Gamma({1\over 2} + i E)} \right]^{1\over 2}, ~~~~~ T_p(E) =  \mu^{iE} \left[ {1\over 1+e^{-2\pi E}}\cdot {\Gamma({1\over 2}-i E)\over \Gamma({1\over 2} + i E)} \right]^{1\over 2}.
\fe
Note that 
\ie
R_h(E) = (R_p(E))^*,~~~~T_h(E)=(T_p(E))^*
\fe
have the interpretation of the reflection and transmission coefficient of a hole in the fermi sea (at energy $E<-\mu$).\footnote{In contrast, in the $c=1$ matrix quantum mechanics, due to a difference choice of the fermi sea ``ground" state $|\Omega\rangle$, the ``hole" reflection coefficient is instead $R_h(E) = \left( R_p(E) \right)^{-1}$\cite{Balthazar:2017mxh}. }

By definition, $[\varphi_E^{(L)}]_R^\leftarrow=0$, and thus $\varphi_E^{(L)}$ do not appear in (\ref{lszaoo}). The sum over fermion states only involves those of the form $\varphi_E^{(R)}$. Taking the continuum limit of (\ref{lszaoo}) results in the formula
\ie\label{aeer}
{\cal A}_{1_R\to 1_R} (\omega) &= \int {dE\over 2\pi} \int {dE' \over 2\pi} \int_{-\infty}^\infty dz\, {R_p(E)\over z -E+i\epsilon(E+\mu)} {R_h(E') \over z-\omega - E' + i\epsilon(E'-\mu)}.
\fe
Note that $R_p(E)$ is an analytic function on the upper half complex $E$-plane (all of its poles lie on the lower half plane), as is generally expected of an S-matrix element in non-relativistic quantum mechanics. We cam thus deform the integration contour in $E$ to $\mathbb{R}+i\infty$, picking up residues at
\ie\label{residuee}
E_* = {z + i\epsilon\mu\over 1-i\epsilon}
\fe
for $z>-\mu$ so that ${\rm Im}E_*>0$. Likewise, we can deform the integration contour in $E'$ to $\mathbb{R}-i\infty$, picking up residues at 
\ie\label{residueepr}
E_*' = {z-\omega + i\epsilon\mu \over 1-i\epsilon}
\fe
for $z<\omega-\mu$ so that ${\rm Im} E_*'<0$. Thus we arrive at 
\ie\label{aoor}
{\cal A}_{1_R\to 1_R} (\omega) &= \int_{-\mu}^{\omega-\mu} dz \, R_p(z) R_h(z-\omega).
\fe
This derivation can be generalized to in- and out-states that involve an arbitrary of boson/collective modes, in both left and right asymptotic regions. The resulting amplitudes can always be written as an integral of product of fermion reflection or transmission coefficients \cite{Moore:1991zv}.

\subsection{$1\to k$ amplitudes in the uncharged sector}
\label{sec:unchargedmqmamp}

\subsubsection{Reflection of a particle-hole pair}

Following the notation for S-matrix elements in section 2.2 of \cite{Balthazar:2022atu}, the amplitude of a single collective mode reflecting off the barrier into $k$ collective modes of energies $\omega_1,\ldots,\omega_k$ on the \emph{same} side of the fermi sea is given by
\ie
\cA_{1_{R/L}\to k_{R/L}}(\omega_1,\ldots,\omega_k)
&= - \sum_{S_1\sqcup S_2=S}(-1)^{|S_2|}\int_0^{\omega(S_2)}dx \, R_p(-\mu+\omega-x) R_h(-\mu-x).
\label{eq:mmRR}
\fe
Here $S_1$ and $S_2$ are disjoint subsets of $S = \{\omega_1, \ldots, \omega_k\}$ such that $S_1\sqcup S_2=S$. $|S_2|$ denotes the number of elements of $S_2$, and $\omega(S_2)$ the sum of all elements of $S_2$. 
The integrand of (\ref{eq:mmRR}) can be written as
\ie
\left[ \frac{1}{1+e^{-2\pi\mu}e^{2\pi(\omega-x)}}\frac{1}{1+e^{-2\pi\mu}e^{-2\pi x}} \right]^{1\over 2} K(\mu,\omega,x), 
\label{eq:mmRRintegrand}
\fe
where
\ie
K(\mu, \omega, x) &\equiv \mu^{-i\omega} \left[ {\Gamma({1\over 2} - i (-\mu+\omega-x)) \over \Gamma({1\over 2} - i (-\mu - x))} {\Gamma({1\over 2} + i (-\mu-x))\over \Gamma({1\over 2} + i(-\mu+\omega-x))} \right]^{1\over 2}.
\label{eq:Kfn}
\fe
The asymptotic series expansion of $K(\mu, \omega, x)$ in $\mu^{-1}$, integrated in $x$ according to (\ref{eq:mmRR}), gives the perturbative expansion of the amplitude. Its agreement with with type 0B closed string perturbation theory has been analyzed in \cite{Balthazar:2022atu}. Note that the perturbative expansion of (\ref{eq:mmRR}) is identical to that of the $1\to k$ amplitude in the $c=1$ MQM. On the other hand, the prefactor multiplying $K(\mu, \omega, x)$ in (\ref{eq:mmRRintegrand}) gives rise to non-perturbative corrections to the amplitude, starting at order $e^{-2\pi\mu}$, which are distinct from that of the $c=1$ MQM \cite{Balthazar:2017mxh}.

Explicitly, $\cA_{1_{R/L}\to k_{R/L}}$ admits an expansion of the form
\ie
\cA_{1_{R/L}\to k_{R/L}}(\omega_1,\ldots,\omega_k)&= \sum_{g=0}^{\infty}\frac{1}{\mu^{k-1+2g}}{\cal A}^{\text{pert},(g)}_{1_{R/L}\to k_{R/L}}(\omega_1,\ldots,\omega_k)\\
&~~~ +\sum_{n=1}^\infty e^{-2\pi n \mu}\sum_{L=0}^\infty \frac{1}{\mu^{{L}}}{\cal A}^{2n-\text{inst},(L)}_{1_{R/L}\to k_{R/L}}(\omega_1,\ldots,\omega_k),
\label{eq:mmRRinstexpansion}
\fe 
where the power series in $\mu^{-1}$ appearing in both terms on the RHS are asymptotic and Borel summable, and the summation over $g$ and $L$ are understood as defined via Borel resummation \cite{Balthazar:2019rnh}.

The first sum on the RHS of (\ref{eq:mmRRinstexpansion}) is to be identified with the closed string genus expansion. In the second line of (\ref{eq:mmRRinstexpansion}), the summation over $n$ amounts to the D-instanton expansion, and the summation over $L$ is to be identified with the open string loop expansion in a given D-instanton sector. For instance, the leading non-perturbative correction at order $e^{-2\pi\mu}$ is
\ie
{\cal A}^{2-\text{inst},(0)}_{1_{R/L}\to k_{R/L}}(\omega_1, \ldots , \omega_k)= -\frac{2^{k-1}}{\pi}\cosh\Big(\pi\sum_{j=1}^k\omega_j\Big)\prod_{i=1}^k\sinh(\pi\omega_i),
\label{eq:mmRR1tokleading}
\fe


\subsubsection{Transmission of a particle-hole pair}

The amplitude of a collective mode tunneling through the potential barrier, and turning into $k$ collective modes on the \emph{opposite} side of the fermi sea of energies $\omega_1,\ldots,\omega_k$, is given by
\ie
\cA_{1_{R/L}\to k_{L/R}}(\omega_1,\ldots,\omega_k) 
&= - \sum_{S_1\sqcup S_2=S}(-1)^{|S_2|}\int_0^{\omega(S_2)}dx \, T_p(-\mu+\omega-x) T_h(-\mu-x).
\label{eq:mmTT}
\fe
The integrand of (\ref{eq:mmTT}) can be written as
\ie
\left[ \frac{1}{1+e^{2\pi\mu}e^{-2\pi(\omega-x)}}\frac{1}{1+e^{2\pi\mu}e^{2\pi x}} \right]^{1\over 2} K(\mu,\omega,x),
\label{eq:mmTTintegrand}
\fe
where $K$ is defined as in (\ref{eq:Kfn}). 
The expansion for $\cA_{1_{R/L}\to k_{L/R}}$ analogous to (\ref{eq:mmRRinstexpansion}) takes the form
\ie
\cA_{1_{R/L}\to k_{L/R}}(\omega_1,\ldots,\omega_k)= \sum_{n=1}^\infty e^{-2\pi n \mu}\sum_{L=0}^\infty \frac{1}{\mu^{{L}}}{\cal A}^{2n-\text{inst},(L)}_{1_{R/L}\to k_{L/R}}(\omega_1,\ldots,\omega_k).
\label{eq:mmTTinstexpansion}
\fe
The leading term at order $e^{-2\pi\mu}$ is given by 
\ie
{\cal A}^{2-\text{inst},(0)}_{1_{R/L}\to k_{L/R}}(\omega_1, \ldots , \omega_k)= \frac{2^{k-1}}{\pi}\prod_{i=1}^k\sinh(\pi\omega_i).
\label{eq:mmTT1tokleading}
\fe

\subsection{Finite temperature}
\label{sec:finittmqm}

An alternative derivation of formulae such as (\ref{aoor}) makes use of the bosonization relation
\ie\label{abbbos}
a_{L/R}(\omega) = \int_{-\infty}^\infty dy \, b_{L/R}^\dagger(-\mu+y) b_{L/R}(-\mu+y - \omega),
\fe
where $a_{L/R}$ are the boson annihilation operators in the left/right asymptotic regions of the fermi sea, and $b_{L/R}$ analogous for the fermions. This allows for a straightforward generalization of (\ref{aoor}) to the thermal S-matrix element
\ie
{\cal S}^{(T)}_{1_R\to 1_R}(\omega'|\omega) = \delta(\omega'-\omega) {\cal A}^{(T)}_{1_R \to 1_R}(\omega),
\fe
with
\ie\label{atomeg}
{\cal A}^{(T)}_{1_R \to 1_R}(\omega) = \int_{-\infty}^\infty dy \,{R_p(-\mu+y) R_h(-\mu+y-\omega) \over (1+e^{-2\pi R y}) (1+e^{-2\pi R(\omega-y)})},
\fe
where the denominators in the integrand come from the Fermi-Dirac distribution of the fermions in the grand canonical ensemble of temperature $T={1\over 2\pi R}$ and chemical potential $-\mu$.

To make contact with Euclidean thermal Green functions, we consider the analytic continuation of (\ref{atomeg}) by moving $\omega$ into the upper half complex plane toward the Matsubara frequency ${in\over R}$, where $n\in \mathbb{Z}_{>0}$. In this procedure, the integrand on the RHS of (\ref{atomeg}) has $n$ poles at
\ie
y = \omega - {i(m+{1\over 2})\over R},~~~~m=0,\ldots,n-1,
\fe
that move across the real $y$-axis, giving the residue contribution
\ie
{i\over R} {1\over 1-e^{-2\pi R \omega}} \sum_{m=0}^{n-1} R_p(-\mu +\omega - {i(m+{1\over 2})\over R}) R_h(-\mu - {i(m+{1\over 2})\over R}).
\fe
Note that the Bose-Einstein factor that arises from one of the denominators in (\ref{atomeg}) at the pole is divergent in the limit $\omega \to {in\over R}$.
The remaining $y$-integral is finite, as there are no singularities of $R_h(-\mu+y-{in\over R})$ at real $y$ and positive $n$. Thus we find in the $\omega \to {in\over R}$ limit,
\ie\label{attome}
{\cal A}^{(T)}_{1_R \to 1_R}(\omega) \to {1\over 1-e^{-2\pi R \omega}} \widetilde {\cal A}^{(T)}_{1_R \to 1_R}(\omega={in\over R}),
\fe
where 
\ie\label{tildeattaoo}
\widetilde {\cal A}^{(T)}_{1_R \to 1_R}(\omega={in\over R}) = {i\over R} \sum_{m=0}^{n-1} R_p(-\mu + {in\over R} - {i(m+{1\over 2})\over R}) R_h(-\mu - {i(m+{1\over 2})\over R}).
\fe

(\ref{attome}) is a special case of the property of the thermal S-matrix element (\ref{finitetsdef}) upon analytic continuation of the energies to Matsubara frequencies, as follows. Firstly, the thermal S-matrix elements of the collective bosons of the MQM can be written as cluster sums of the form
\ie
{\cal S}^{(T)}(\B|\A) = \sum_{\{\B|\A\}=\sqcup \{\B_I,\A_I\}} \prod_I {\cal S}^{(T)}_{{\rm conn}}
(\B_I|\A_I),
\fe
where each connected thermal S-matrix element ${\cal S}^{(T)}_{{\rm conn}}(\B_I|\A_I)$ involves a single fermion loop, similarly to the ordinary S-matrix computed in \cite{Moore:1991zv}. We have
\ie
{\cal S}^{(T)}_{{\rm conn}}(\B|\A) = \delta(E_\B-E_\A) {\cal A}^{(T)}(\B|\A),
\fe
where ${\cal A}^{(T)}(\B|\A)$ is an analytic function of the energies of the asymptotic particles subject to energy conservation. In the ``Matsubara limit", which we define as taking each of the independent boson energies to a Matsubara frequency, ${\cal A}^{(T)}(\B|\A)$ is given by the product of a set of divergent Bose-Einstein statistical factors and a finite ``thermal amplitude" $\widetilde{\cal A}^{(T)}(\B|\A)$ evaluated at the Matsubara frequencies.
For instance, a $1\to k$ thermal S-matrix element ${\cal S}^{(T)}_{1\to k}(\omega_1,\ldots,\omega_k|\omega) = \delta(\omega-\sum_{j=1}^k\omega_j) {\cal A}^{(T)}(\omega_1,\ldots,\omega_k)$ obeys
\ie\label{attildeaa}
{\cal A}^{(T)}(\omega_1,\ldots,\omega_k) \to \prod_{j=1}^k{1\over 1-e^{-2\pi R\omega_j}}\cdot  \widetilde{\cal A}^{(T)}(\omega_1,\ldots,\omega_k)
\fe
in the limit $\omega_j \to {i n_j\over R}$, $n_j\in \mathbb{Z}_{>0}$. We refer to (\ref{attildeaa}) as a ``thermal LSZ" relation.

The thermal amplitude $\widetilde{\cal A}^{(T)}$ can also be calculated directly from an LSZ-like limit of Euclidean thermal Green function, in a way similar to the derivation of section \ref{sec:lszred}. Using the thermal free fermion propagator
\ie
{\rm Tr} \left[ \rho^{(T)} {\bf T} \psi(x_1, -i\tau_1) \psi^\dagger(x_2, -i\tau_2) \right] =  -{1\over 2\pi R} \sum_n \varphi_n(x_1) \varphi_n^*(x_2) \sum_{r\in \mathbb{Z}+{1\over 2}} {e^{- (-\mu + {ir\over R})  \tau_{12}} \over -\mu+ {i r\over R} - E_n + i\epsilon(E_n+\mu)},
\fe
we can extract for instance $\widetilde{\cal A}^{(T)}_{1_R\to 1_R}$ with a formula analogous to (\ref{aeer}), 
\ie\label{matsua}
{\cal A}_{1_R\to 1_R}^{(T)} \Big(\omega={in\over R}\Big) &= \int {dE\over 2\pi} \int {dE' \over 2\pi} {i\over R} \sum_{r\in\mathbb{Z}+{1\over 2}} {R_p(E)\over -\mu+{ir\over R} -E+i\epsilon(E+\mu)} {R_h(E') \over -\mu+{ir\over R} -\omega - E' + i\epsilon(E'-\mu)}.
\fe
Evaluating the integrals in $E$ and $E'$ using residues analogous to (\ref{residuee}), (\ref{residueepr}) reproduces (\ref{tildeattaoo}). From the Green function perspective, the Bose-Einstein factor appearing in (\ref{attildeaa}) arises due to the Wick rotation that turns an integration contour at late time along the complexified time cylinder into a spiral contour that densely winds the thermal circle.

It is instructive to inspect the perturbative expansion of (\ref{matsua}). The tree level thermal amplitude is independent of the temperature and is identical to the ordinary tree level scattering amplitude. At genus one, for instance, the $1\to 1$ thermal amplitude is 
\ie\label{genusonetherm}
{\cal A}_{1_R\to 1_R}^{(T),{\rm pert},(1)}(\omega) = -{1\over 24} \omega^2 (\omega-i) \Big( 1 -  i\omega + \omega^2 +{1\over R^2} \Big)
\fe
Such results have been previously obtained in \cite{Klebanov:1991qa}. It would be very interesting to reproduce the temperature dependence in (\ref{genusonetherm}) from a worldsheet computation analogous to the one performed in \cite{Balthazar:2017mxh}.

The formula (\ref{eq:mmRR}) for $1_R\to k_R$ amplitude, for instance, can be extended to the thermal case analogously to (\ref{tildeattaoo}). The leading non-perturbative term in ${\cal A}_{1_R\to k_R}^{(T)}$, at order $e^{-2\pi \mu}$, is simply given by (\ref{eq:mmRR1tokleading}) multiplied by the factor
\ie
{\pi \over R\sin \left({\pi\over R} \right) }.
\label{eq:finTfactor}
\fe

\section{D-instantons in type 0B string theory}
\label{Dinst}

\subsection{ZZ-boundary condition in ${\cal N}=1$ Liouville CFT}

The ${\cal N}=1$ Liouville theory at central charge $c={3\over 2} (1+2Q^2)$, with $Q=b+b^{-1}$, at the particular value of $b=1$ was reviewed extensively in the companion paper \cite{Balthazar:2022atu}. Here we briefly recap the operator content of the theory and its simplest superconformal boundary condition based on the $(1,1)$ degenerate representation of the super-Virasoro algebra.

The operator bulk spectrum in the (NS, NS) sector consists of super-Virasoro primaries $V_P$ labeled by the Liouville momentum $P\in \mathbb{R}_{\geq 0}$, of weight $h=\widetilde h = {1\over 2}(1+P^2)$. 
The (R, R) sector spectrum of ${\cal N}=1$ Liouville theory (viewed as a modular invariant bosonic CFT) consists of lowest weight states $V_P^{{\rm R},+}$ with respect to the superconformal algebra (SCA), of weight $h=\widetilde h = {1\over 16} + {1\over 2}(1+P^2)$, $P\geq 0$. There is a closely related family of defect operators attached to a $\mathbb{Z}_2$ symmetry topological line, which we denote by $V_P^{{\rm R},-}$, and have the same weight $h=\widetilde h = {1\over 16} + {1\over 2}(1+P^2)$.


Similarly to its bosonic cousin, superconformal boundary conditions in $\cN=1$ Liouville theory come in two families, namely of FZZT type and of ZZ type \cite{Fukuda:2002bv}. In the type 0B string context, the FZZT boundary condition defines a semi-infinite brane that extends along the Liouville target space, whose physical open string spectrum consists of a continuum of scattering states. We will be concerned with the boundary condition of ZZ type, which corresponds to a pointlike brane localized in the strong coupling region of the Liouville target space. They come in a discrete family parameterized by a pair of positive integers $m,n$, and are in correspondence with degenerate representations of the $\cN=1$ super-Virasoro algebra; we will refer to them as the $\cN=1$ $(m,n)$ ZZ boundary condition.

In this paper, we will only analyze the effects of the simplest class of possible D-instantons, namely those constructed from the $\cN=1$ $(1,1)$ ZZ boundary condition.\footnote{As was observed in \cite{Balthazar:2019ypi} in the $c=1$ string theory context, the D-instantons that involve higher $(m,n)$ ZZ boundary conditions may contribute at subleading orders in $e^{-1/g_s}$.} The boundary state of the latter takes the form
\ie
\ket{\text{ZZ}} = \int_0^\infty \frac{dP}{\pi} \Psi_{\nssect}(P) |V_P\rangle\rangle + \int_0^\infty \frac{dP}{\pi} \Psi_{\rsect}(P) |V^{\rsect,+}_P\rangle\rangle,
\label{eq:ZZ11bdrystate}
\fe
where $|V_P\rangle\rangle$ is the super-Ishibashi state associated with the primary $V_P$ in the (NS, NS) sector and $|V^{\rsect,+}_P\rangle\rangle$ is the super-Ishibashi state associated with $V^{\rsect,+}_P$ in the (R, R) sector. These Ishibashi states satisfy $\langle\langle V_P| q^{\frac{1}{2}(L_0+\widetilde{L}_0-\frac{c}{12})} |V_{P'} \rangle\rangle = \pi\delta(P-P')\,\tr_{\nssect}\, q^{L_0 - \frac{c}{24}}$ and $\langle\langle V^{\rsect,+}_P| q^{\frac{1}{2}(L_0+\widetilde{L}_0-\frac{c}{12})} |V^{\rsect,+}_{P'} \rangle\rangle = \pi\delta(P-P')\,\tr_{\rsect}\, q^{L_0 - \frac{c}{24}}$, where the traces on the RHS are taken over the corresponding holomorphic super-Virasoro representation (without GSO projection). The anti-ZZ boundary state is related by flipping the sign of the (R, R) sector, 
\ie
\ket{\overline{\text{ZZ}}} = \int_0^\infty \frac{dP}{\pi} \Psi_{\nssect}(P) |V_P\rangle\rangle - \int_0^\infty \frac{dP}{\pi} \Psi_{\rsect}(P) |V^{\rsect,+}_P\rangle\rangle.
\label{eq:ZZ11barbdrystate}
\fe
The boundary operator spectrum of the $(1,1)$ ZZ boundary condition, or equivalently the strip Hilbert space ${\cal H}_o$, consists only of the identity operator and its super-Virasoro descendants subject to the GSO projection.
The modular covariance of the cylinder partition function
\ie
\bra{\text{ZZ}} q'^{\frac{1}{2}(L_0+\widetilde{L}_0- {c\over 12})} \ket{\text{ZZ}} = \tr_{\mathcal{H}_o} \left( \frac{1+(-1)^F}{2} q^{L_0-{c\over 24}} \right)
\fe
where $q=e^{-2\pi t}$ and $q'=e^{-2\pi/t}$, or more explicitly,
\ie
\left( \frac{\theta_3(i/t)}{\eta(i/t)^3} \right)^{\frac{1}{2}}\int_0^\infty \frac{dP}{\pi} \left(\Psi_{\nssect}(P) \right)^2 (q')^{P^2\over 2}
+ \left( \frac{2\theta_2(i/t)}{\eta(i/t)^3} \right)^{\frac{1}{2}}\int_0^\infty \frac{dP}{\pi} \left(\Psi_{\rsect}(P) \right)^2 (q')^{P^2\over 2} \\
= \frac{1}{2} \left[ \left( \frac{\theta_3(it)}{\eta(it)^3} \right)^{\frac{1}{2}}\left(q^{-\frac{1}{2}}-1\right) + \left( \frac{\theta_4(it)}{\eta(it)^3} \right)^{\frac{1}{2}}\left(q^{-\frac{1}{2}}+1\right) \right],
\fe
fixes the disc one-point functions
\ie
\Psi_\nssect(P) &= \sqrt{2\pi} \sinh(\pi P), \\
\Psi_\rsect(P) &= \sqrt{2\pi} \cosh(\pi P).
\label{eq:disk1pt11}
\fe

%
%
%

\subsection{The worldsheet description of D-instantons}

The full worldsheet theory of 2D type 0B string consists of a time-like free boson $X^0$ and its Majorana fermion partner $(\psi^0, \widetilde\psi^0)$, the $\cN=1$ Liouville theory of central charge $c={27\over 2}$, together with the $b,c,\beta,\gamma$ ghost system, subject to a diagonal GSO projection. We refer to section 2 of \cite{Balthazar:2022atu} for a review and the relevant conventions (in particular, we set $\A'=2$). There are two types of closed string states that are massless scalar particles in the asymptotic region, namely the (NS, NS) ``tachyon" and the (R, R) ``axion", whose vertex operators are
\ie
\mathcal{T} ^{\pm}_{\omega} = g_s c\widetilde c \, e^{-\phi-\widetilde{\phi}} e^{\pm i \omega X^0} V_{P=\omega},~~~~~
\cA^{\pm}_{\omega} =  g_s \omega c\widetilde c  \, e^{-{\phi\over 2}-{\widetilde{\phi}\over 2}} e^{\pm i \omega X^0} A^{\pm}_{P=\omega},
\label{eq:vops}
\fe
where $\omega$ is the energy, and the superscript $\pm$ on the LHS refers to in- and out-states respectively. $A_P^\pm$ is an (R, R) operator of the full matter SCFT, related to the ${\cal N}=1$ Liouville operators of the previous subsection by $A^{\pm}_P = \frac{1}{\sqrt{2}}\big(  \sigma^0 V_P^{{\rm R}, +} \pm   \mu^0 V_P^{{\rm R},-} \big)$, where $\sigma^0$ and $\mu^0$ are the spin field and disorder operator in the $(\psi^0,\widetilde\psi^0)$ free fermion CFT. The collective modes of the left and right asymptotic regions of the fermi sea of the type 0B MQM are dual to the vertex operators
\ie
\cL^{\pm}_\omega = \cT^{\pm}_\omega - \cA^{\pm}_\omega, ~~~~~ \cR^{\pm}_\omega = \cT^{\pm}_\omega + \cA^{\pm}_\omega,
\label{eq:LRvertexop}
\fe
modulo the insertion of appropriate picture changing operators.

The elementary D-instanton corresponds to a BRST-invariant boundary condition of the worldsheet theory that is the tensor product of Dirichlet boundary condition in the $(X^0,\psi^0,\widetilde\psi^0)$ CFT localized at $X^0=x^0$, the $(1,1)$ ZZ boundary condition in the $\cN=1$ Liouville theory, and the standard boundary condition in the superconformal ghost system $(b,c,\beta,\gamma)$.\footnote{With the $(X^0,\psi^0,\widetilde\psi^0)$ CFT and ${\cal N}=1$ Liouville theory each viewed as a modular invariant bosonic theory, the total matter CFT is formed as a $\mathbb{Z}_2$ orbifold of their tensor product. The boundary state of the D-instanton involves a linear combination of the boundary states of the unorbifolded matter CFT, without twisted sector contribution. This construction is denoted by $|a\rangle^{\rm D}$ in (1.10) of \cite{Fukusumi:2021zme}.}
Using (\ref{eq:disk1pt11}), the disc 1-point amplitude of a tachyon subject to the D- or $\overline{\rm D}$-instanton boundary condition is given by
\ie
\langle \cT^{\pm}_{\omega} \rangle^{D^2}_{D_{x^0}} = \langle \cT^{\pm}_{\omega} \rangle^{D^2}_{\overline{D}_{x^0}} = \frac{g_s \cdtwo}{2\pi} e^{\pm i\omega x^0} \sqrt{2\pi} \sinh(\pi\omega),
\label{eq:tachyon1pt}
\fe
where we have implicitly included the $c$ ghost insertion that fixes the conformal Killing group. 

The analogous disc 1-point amplitude of an axion is given by
\ie
\langle \cA^{+}_{\omega} \rangle^{D^2}_{D_{x^0}} = - \langle \cA^{+}_{\omega} \rangle^{D^2}_{\overline{D}_{x^0}} &= \frac{g_s \cdtwo}{2\pi} e^{i\omega x^0} \sqrt{2\pi} \cosh(\pi\omega),\\
\langle \cA^{-}_{\omega} \rangle^{D^2}_{D_{x^0}} = - \langle \cA^{-}_{\omega} \rangle^{D^2}_{\overline{D}_{x^0}} &= - \frac{g_s \cdtwo}{2\pi} e^{-i\omega x^0} \sqrt{2\pi} \cosh(\pi\omega).
\label{eq:axion1pt}
\fe
These formulae require some explanation, as the axion vertex operator ${\cal A}_\omega^\pm$ is a priori defined in the $(-{1\over 2},-{1\over 2})$ picture, while the disc topology requires total picture number $-2$. Instead, the disc amplitudes (\ref{eq:axion1pt}) are defined indirectly through the disc open-closed 2-point amplitude where an additional open string vertex operator $c e^{-\phi} \psi^{0}$ that corresponds to the D-instanton collective coordinate\footnote{Its picture-raised version is proportional to $\partial X^0$.} is inserted. This open-closed amplitude is related to (\ref{eq:axion1pt}) by differentiating with respect to the collective coordinate $x$. In particular, this leads to the relative minus sign between the in- and out- axion amplitudes in (\ref{eq:axion1pt}).



The RR (axion) charge of the D-instanton is proportional to the zero-momentum limit of the axion disc one-point amplitude. Our convention will be that the D-instanton has  charge $+1$ and the $\overline{\rm D}$-instanton has charge $-1$. The MQM interpretation of the D-instanton is the tunneling, or ``half bounce" in Euclidean time, of a fermion through the potential barrier as depicted in Figure \ref{fig:DinstTunneling}. Thus, we expect the D-instanton action to be equal to that of the half bounce,\footnote{In contrast, the ZZ-instanton of the bosonic $c=1$ string theory corresponds to a full bounce, across and back through the potential barrier, whose action is $2\pi\mu$ \cite{Balthazar:2019rnh}. }
\ie
S_{D} = \pi\mu. 
\fe
The precise relation between $\mu$ and the string coupling $g_s$ is determined in \cite{Balthazar:2022atu} to be $g_s = \frac{4}{\pi\mu}$, and thus $S_{D} = \frac{4}{g_s}$.

\subsection{D-$\overline{\rm D}$ instanton measure and contour}

The leading uncharged instanton contribution to the reflection amplitude of closed strings comes from a D-$\overline{\rm D}$ pair. In this case there are two collective coordinates $x_1^0$ and $x_2^0$ parameterizing the loci of the D- and $\overline{\rm D}$-instanton in time, and the measure on the instanton moduli space depends nontrivially on the separation $x_1^0-x_2^0 = -i \Delta x$. This dependence is expected to be captured by the exponentiated cylinder diagram between the D and $\overline{\rm D}$ boundary conditions \cite{Balthazar:2019ypi},
\ie
\tikz[baseline={([yshift=-1.6ex]current bounding box.center)}]{\pic{cylnnrb0={1}{2}}} &=  \int_0^\infty \frac{dt}{2t} \frac{1}{2}\left[ \left( \frac{\theta_3(it)}{\eta(it)^3} \right)^{1\over 2} \left( q^{- \frac{1}{2}} - 1 \right) \left( \frac{\theta_3(it)}{\eta(it)} \right)^{1\over 2} \frac{e^{-\left(\frac{\Delta x}{2\pi}\right)^2 \pi t}}{\eta(it)} \frac{\eta(it)}{\theta_3(it)} \eta(it)^2\right. \\
& ~~~~~~~~~~~~~~ +\left. \left( \frac{\theta_4(it)}{\eta(it)^3} \right)^{1\over 2} \left( q^{- \frac{1}{2}} + 1 \right) \left( \frac{\theta_4(it)}{\eta(it)} \right)^{1\over 2} \frac{e^{-\left(\frac{\Delta x}{2\pi}\right)^2 \pi t}}{\eta(it)} \frac{\eta(it)}{\theta_4(it)} \eta(it)^2\right] \\
&= \int_0^\infty \frac{dt}{2t} \frac{1}{2}\left[ \left( q^{-\frac{1}{2}} - 1 \right) + \left( q^{-\frac{1}{2}} + 1 \right) \right] e^{-\left(\frac{\Delta x}{2\pi}\right)^2 \pi t},
\label{eq:annulusnaive}
\fe
where $q=e^{-2\pi t}$. Here $t$ is the modulus of the cylinder, the factor ${1\over 2t}$ is due to gauge fixing the conformal Killing group, and the rest of the integrand comes from the partition function of open string states, subject to GSO projection and with the appropriate ghost insertions. In the two terms appearing in the square bracket, we have included the contribution from the ${\cal N}=1$ Liouville, the free fermion and boson, the $\B,\C$, and the $b,c$ ghost system, in that order, without and with $(-)^F$ insertion.


Note that the integral in (\ref{eq:annulusnaive}) is divergent as $t\to 0$, regardless of the separation in Euclidean time $\Delta x$. In section \ref{sec:finiteT}, we will treat this divergence with a careful regularization at finite temperature, namely replacing the Wick rotated free boson $X$ by a compact boson at radius $R$. We will see that even after the regularization, the diagram (\ref{eq:annulusnaive}) still contains a divergence in the zero temperature limit $R\to \infty$, but the latter cancels against that of the cylinder diagram whose boundaries lie on the same D- or $\overline{\rm D}$-instanton, giving a finite total result for the measure factor
\ie
\exp \Big( \tikz[baseline={([yshift=-1.6ex]current bounding box.center)}]{\pic{cylnnrr0={1}{1}}} + \tikz[baseline={([yshift=-1.6ex]current bounding box.center)}]{\pic{cylnnbb0={2}{2}}} + 2~\tikz[baseline={([yshift=-1.6ex]current bounding box.center)}]{\pic{cylnnrb0={1}{2}}} \Big) = {\cN_D^2 \over \Delta x^2 - (2\pi)^2}.
\label{eq:measure}
\fe
Here $\cN_D$ is a normalization constant that cannot be determined a priori in the worldsheet formalism.\footnote{They are analogous to the normalization constants $\cN_m$ associated with $(m,1)$ ZZ instantons in $c=1$ string theory \cite{Balthazar:2019ypi}.} Instead, we will fix them by a direct comparison against the anticipated dual MQM results presented in section \ref{sec:mqmreview}.

It is interesting to compare the integration measure (\ref{eq:measure}) on D-$\overline{\rm D}$ instanton moduli space in type 0B string theory with the integration measure on multi-ZZ-instanton moduli space in $c=1$ string theory \cite{Balthazar:2019ypi} . 
The logarithm of the D-$\overline{\rm D}$ measure (\ref{eq:measure}) is proportional to $\log \frac{1}{\Delta x^2 - (2\pi)^2}$, which in the limit of large Euclidean time separation $\Delta x \to \infty$ behaves as $-\log \Delta x^2$. The latter is analogous to the RR Coulomb potential that arises from the cylinder diagram between a pair of charged D-branes \cite{Polchinski:1995mt}.
In contrast, the logarithm of the measure factor in the separation of a pair of ZZ-instantons in $c=1$ string theory is proportional to $\log \frac{\Delta x^2}{\Delta x^2 - (2\pi)^2}$, which vanishes in the $\Delta x\to \infty$ limit. 

\begin{figure}[h!]
\centering
\begin{tikzpicture}
[decoration={markings,mark=at position 0.6 with {\arrow{>}}}] 
\draw[color=gray] (-3,0) -- (3,0);
\draw[color=gray] (0,-2.3) -- (0,2.3);
\draw (-1,0) node[cross=4pt, very thick] {};
\node at (-1,-0.5) {$-2\pi$};
\draw (1,0) node[cross=4pt, very thick] {};
\node at (1,-0.5) {$2\pi$};
\draw[thick] (-3,0) -- (-1.2,0);
\draw[thick,postaction={decorate}] (-0.8,0) -- (0.8,0);
\draw[thick] (1.2,0) -- (3,0);
\draw (-2.2,0) node[above] {$\mathcal{C}$};
\node(n)[inner sep=2pt] at (-3,2) {$\Delta x$};
\draw[line cap=round](n.south west)--(n.south east)--(n.north east);
\end{tikzpicture}
\caption{The integration over the D-$\overline{\rm D}$ instanton moduli space is defined using the principal value contour along Euclidean time. 
}
\label{fig:contourPV}
\end{figure}
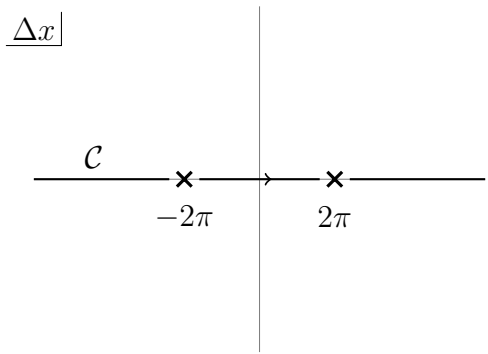

The last piece of ingredient in the worldsheet formulation of D-instanton is a choice of contour prescription in the integration over the instanton moduli space, due to the poles appearing in (\ref{eq:measure}). The latter have to do with the appearance of additional on-shell open string modes, when the D- and $\overline{\rm D}$-instanton are separated at $\Delta x=\pm 2\pi$. Our proposal is to perform the moduli integration with a {\it principal value} contour, as shown in Figure \ref{fig:contourPV}. In section \ref{sec:dinstuncharged}, we will find that the principal value prescription leads to amplitudes that agree with the dual MQM results in a highly nontrivial manner. Note that this is different from the contour prescription for integrating multiple ZZ-instantons in $c=1$ string theory adopted in \cite{Balthazar:2019ypi}.

In a string field theoretic approach to D-instantons, the integration over the instanton moduli space is part of the functional integration over open string fields, where a contour prescription is chosen in the open string field space. It was argued in \cite{Sen:2020ruy} that a ``unitary contour" prescription is required in order for the closed string (field) amplitudes with D-instanton corrections to obey Cutkowsky cutting rules, extending that of perturbative unitarity. This leads to results in agreement with the principal value contour in the on-shell formulation considered here. This is perhaps unsurprising as the closed string amplitudes in type 0B string theory are expected to be unitary (provided that soft quanta are taken into account appropriately in the charged processes), in contrast to $c=1$ string theory where the closed string sector is not unitary on its own.


\subsection{D-instantons at finite temperature}
\label{sec:finiteT}

Type 0B string theory at finite temperature $T=\frac{1}{2\pi R}$ is described on the worldsheet as Wick rotating the free boson $X^0=-i X$, and compactifying $X$ to the thermal circle
\ie
X \sim X + 2\pi R.
\fe
This leads to the restriction that closed string modes have integer units of ``momentum", or rather Matsubara frequencies $\omega = {in\over R}$, $n\in\mathbb{Z}$, as well as integer winding numbers around the thermal circle. In this paper we will only consider closed string amplitudes involving the ``momentum" modes, which we propose to be dual to thermal amplitudes at Matsubara frequencies defined via the ``thermal LSZ" limit of section \ref{sec:finittmqm}.\footnote{See \cite{Jafferis:2021ywg} however for interesting interpretations of the closed string winding modes on the thermal circle in a closely related context.}

The open string modes on the D-instanton, on the other hand, can acquire nontrivial winding number $w$ around the thermal circle. In particular, the cylinder diagram with both boundary components on the same D-instanton is computed by
\ie
\label{eq:DDfinT}
\tikz[baseline={([yshift=-1.6ex]current bounding box.center)}]{\pic{cylnnrr0={}{}}}
&=\int_0^\infty \frac{dt}{2t} \frac{1}{2}\left[ \left( \frac{\theta_3(it)}{\eta(it)^3} \right)^{1\over 2} \left( q^{- \frac{1}{2}} - 1 \right) \left( \frac{\theta_3(it)}{\eta(it)^3} \right)^{1\over 2} \frac{\eta(it)}{\theta_3(it)} \eta(it)^2\right. \\
& ~~~~~~~~~~~~~~ -\left. \left( \frac{\theta_4(it)}{\eta(it)^3} \right)^{1\over 2} \left( q^{- \frac{1}{2}} + 1 \right) \left( \frac{\theta_4(it)}{\eta(it)^3} \right)^{1\over 2} \frac{\eta(it)}{\theta_4(it)} \eta(it)^2\right]  \sum_{w\in\bZ} e^{-(wR)^2 \pi t} \\
&=-\int_0^\infty \frac{dt}{2t}  \sum_{w\in\bZ} e^{-(wR)^2 \pi t}.
\fe
Note that the two terms appearing in the square bracket are the same as those in (\ref{eq:annulusnaive}) with $\Delta x$ set to zero, and the relative sign between them flipped. The latter is due to the GSO projection on D-D open string modes, which is different from that of the D-$\overline{\rm D}$ open string modes in (\ref{eq:annulusnaive}). Furthermore, (\ref{eq:DDfinT}) includes a sum over the winding number $w$.

The integral in \eqref{eq:DDfinT} is divergent near the boundary of the moduli space of the cylinder geometry. Near $t=0$, the integrand behaves as $t^{-{3\over 2}}$, which gives rise to a power divergence. As is familiar in string perturbation theory, such power divergence can be consistently regularized by a simple subtraction. There is also a logarithmic divergence near $t=\infty$ that comes entirely from the zero winding ($w=0$) sector, which we regularize by cutting off the cylinder moduli space at $t\leq L$, and take $L\to \infty$ after including a counter term. Note that the counter term for canceling the log divergence is independent of $R$, or the temperature. The resulting regularized cylinder amplitude is
\ie\label{eq:cylDD}
\tikz[baseline={([yshift=-1.6ex]current bounding box.center)}]{\pic{cylnnrr0={}{}}} 
&= \lim_{L\to \infty} \left\{ {1\over 2} \log {L\over L_D} + \int_0^L dt \bigg[ -\frac{1}{2t} \sum_{w\in\bZ} e^{-(wR)^2 \pi t} + \frac{1}{2R} t^{-\frac{3}{2}} \bigg] \right\}
\\
&= \log{{\cal N}_D\over R}.
\fe
Here $L_D$ or ${\cal N}_D$ is a finite ($R$-independent) constant that is a priori undetermined in the worldsheet formalism.\footnote{It may be possible to derive this constant from string field theory, as was done in \cite{Sen:2021qdk} for the $(1,1)$ ZZ-instanton in $c=1$ string theory.} The same formula holds for the cylinder amplitude with boundaries on an $\overline{\rm D}$-instanton. 

An immediate consequence of the calculation (\ref{eq:cylDD}) is that the measure factor for a single D-instanton is proportional to ${\cal N}_D/R$, which vanishes in the zero temperature ($R\to \infty$) limit. This is in agreement with the expectation from the dual MQM that the charged sector processes are IR divergent, and in particular the exclusive amplitudes for charged process with respect to Fock basis asymptotic states vanish.

Similarly, the cylinder diagram between a D-instanton and a $\overline{\rm D}$-instanton, separated by $\Delta x$ in Euclidean time, is computed by
\ie\label{eq:cylDDbar}
\tikz[baseline={([yshift=-1.6ex]current bounding box.center)}]{\pic{cylnnrb0={1}{2}}} 
&= \int_0^\infty dt \bigg[ \frac{e^{\pi t}}{2t} \sum_{w\in\bZ} e^{-(wR+\frac{\Delta x}{2\pi})^2 \pi t} - \frac{1}{2R} t^{-\frac{3}{2}}  \bigg] 
\\
&= -{1\over 2} \log \left[ 4\sin\left( \frac{\Delta x -2\pi}{2R} \right)\sin\left( \frac{\Delta x + 2\pi}{2R} \right) \right],
\fe
where we have included the counter term that subtracts the power divergence at $t=0$. Note that there is no divergence from the $t\to\infty$ limit for sufficiently large $R$ and a suitable range of $\Delta x$, and we extend the result to all values of $R$ and $\Delta x$ by analytic continuation.

Altogether, the measure factor on the D-$\overline{\rm D}$ instanton moduli space due to the cylinder diagrams is
\ie
\exp \Big( \tikz[baseline={([yshift=-1.6ex]current bounding box.center)}]{\pic{cylnnrr0={1}{1}}} + \tikz[baseline={([yshift=-1.6ex]current bounding box.center)}]{\pic{cylnnbb0={2}{2}}} + 2~\tikz[baseline={([yshift=-1.6ex]current bounding box.center)}]{\pic{cylnnrb0={1}{2}}} \Big) = {\cN_D^2 \over 4 R^2 \sin\left( \frac{\Delta x -2\pi}{2R} \right)\sin\left( \frac{\Delta x + 2\pi}{2R} \right)},
\label{eq:DDbarfinT}
\fe
In the limit $R\to \infty$, it reduces to (\ref{eq:measure}) as claimed.



\section{D-instanton mediated amplitudes}
\label{sec:dinstuncharged}


In this section we compute the leading nontrivial D-instanton contribution to the uncharged $1_R\to k_R$ and $1_R\to k_L$ closed string amplitudes. The notation here follows that of section \ref{sec:mqmreview}, where the subscript $L$ and $R$ represent closed strings dual to collective modes in the left and right asymptotic regions of the fermi sea, represented on the worldsheet by the vertex operators (\ref{eq:LRvertexop}).

We follow the same convention as \cite{Balthazar:2022atu}, in which the closed string energy $\omega_{\rm WS}$ appearing in the vertex operators in the worldsheet description (e.g. $\omega$ in (\ref{eq:tachyon1pt}) and (\ref{eq:axion1pt})) differs from the energy $\omega_{\rm MM}$ of the collective mode in the dual matrix model (e.g. $\omega$ in (\ref{aoor})) by a factor of 2, namely
\ie
\omega_{\text{MM}} = 2\,\omega_{\rm WS}.
\fe
Furthermore, as we have stripped off the energy conservation delta function $\delta(E_\B-E_\A)$ from the S-matrix element in the definition of scattering amplitudes, the amplitude ${\cal A}$ in the MQM convention and $\mathfrak{A}$ in the string worldsheet convention are related by an overall factor of $2$ due to the rescaling of energy, namely
\ie\label{mmwszerot}
\mathfrak{A}(\{\omega_\B\}, \{\omega_\A\}) = {1\over 2} {\cal A}(\{ 2\omega_\B\}, \{ 2\omega_\A\}) .
\fe
At finite temperature $T={1\over 2\pi R}$, the energies of the worldsheet vertex operators are restricted to Matsubara frequencies, namely $\omega={in\over R}$ for $n\in\mathbb{Z}$, and the string amplitude $\mathfrak{A}^{(T)}$ is defined through the worldsheet computation with compactified target Euclidean time, with an overall factor of $(-iR)$ stripped off. Our proposal of the matrix model dual is the thermal amplitude $\widetilde{\cal A}$ defined through the ``thermal LSZ" limit (\ref{attildeaa}), related to the string amplitude by
\ie\label{mmwsfinitet}
\mathfrak{A}^{(T)}(\{\omega_\B\}, \{\omega_\A\}) = {1\over 2} \widetilde{\cal A}^{(2T)}(\{ 2\omega_\B\}, \{ 2\omega_\A\}) .
\fe

In section \ref{chargedfinitews}, we will also consider a finite temperature string amplitude mediated by a single charged D-instanton, but their precise MQM interpretation as a sort of thermal amplitude for charged processes remains to be clarified.

\subsection{D-$\overline{\rm D}$ instanton contribution to the uncharged $1_R\to k_R$ amplitude}

The leading non-perturbative contribution to the uncharged $1_R\to k_R$ closed string amplitude is already a multi-instanton process, mediated by a D-$\overline{\rm D}$ instanton pair. Similarly to the instanton calculus of \cite{Balthazar:2019ypi}, the corresponding S-matrix element is calculated by integrating a product of disc diagrams over the D-instanton moduli space,
\ie
e^{-2 S_{D}} \cN_{D}^2 \int  \frac{dx_1dx_2}{\left(\Delta x\right)^2 - (2\pi)^2} \biggl( \tikz[baseline={([yshift=-1.6ex]current bounding box.center)}]{\pic{disk1r={1}}} \,+\, \tikz[baseline={([yshift=-1.6ex]current bounding box.center)}]{\pic{disk1b={2}}} \biggr) \biggl( \tikz[baseline={([yshift=-1.6ex]current bounding box.center)}]{\pic{disk1r={1}}} \,+\, \tikz[baseline={([yshift=-1.6ex]current bounding box.center)}]{\pic{disk1b={2}}} \biggr) \cdots \biggl( \tikz[baseline={([yshift=-1.6ex]current bounding box.center)}]{\pic{disk1r={1}}} \,+\, \tikz[baseline={([yshift=-1.6ex]current bounding box.center)}]{\pic{disk1b={2}}} \biggr).
\label{eq:DDbar11}
\fe
where $x_1, x_2$ are Euclidean time coordinates of the D- and $\overline{\rm D}$-instanton, and $\Delta x = x_1 -x_2$. Here we have included the measure factor (\ref{eq:measure}) computed from the cylinder diagrams. Each factor of the product in the integrand of (\ref{eq:DDbar11}) is the sum of a pair of disc one-point amplitudes with a D- or $\overline{\rm D}$-instanton boundary condition. On the first pair of discs, we insert the in-state vertex operator $\cR^{+}_\omega$, while each of the remaining $k$ pairs of discs comes with the insertion of an out-state vertex operator $\cR^{-}_{\omega_i}$, $i=1,\dots,k$. In total, the product of the discs gives
\ie{}
\left( g_s \cdtwo \over \sqrt{2\pi} \right)^{k+1} \left( e^{i\pi\omega^E}e^{i \omega^E x_1} - e^{i\omega^E x_2}e^{-i\pi\omega^E} \right) \prod_{i=1}^k \left( -e^{-i\pi\omega^E_{i}}e^{-i\omega^E_i x_1} + e^{-i\omega^E_{i}x_2}e^{i\pi\omega^E_{i}} \right),
\label{eq:DDbar1RtokRstep1}
\fe
where $\omega_j^E$ are related to the energies $\omega_j$ of the out-modes by $\omega_j = i \omega_j^E$.

The integral over the overall ``center of mass" coordinate of the D-$\overline{\rm D}$ instanton pair in (\ref{eq:DDbar11}) yields a factor $2\pi i\, \delta(\omega - \sum_{i=1}^k\omega_k)$, where the $i$ comes from the Wick rotation of time contour. After factoring out the energy-conservation delta function, we are left with the amplitude expressed as an integral over the D-instanton separation $\Delta x$, which we perform for imaginary energies (i.e. real $\omega_j^E$) using the principal value contour prescription as shown in Figure \ref{fig:contourPV},
\ie\label{eq:DDbar1RtokRstep2}
\mathfrak{A}^{D\overline D}_{1_R\to k_R}(\omega_1,\ldots,\omega_k)
&=  2\pi i \, e^{-2 S_{D}} \cN_{D}^2 \left( g_s \cdtwo \over \sqrt{2\pi} \right)^{k+1} \int_{{\cal C}} \frac{d\Delta x}{(\Delta x)^2-(2\pi)^2} \left( e^{i\pi\omega^E} - e^{-i\omega^E \Delta x}e^{-i\pi\omega^E} \right) 
\\
&~~~~\times\prod_{i=1}^k \left( -e^{-i\pi\omega^E_{i}} + e^{i\omega^E_{i}\Delta x}e^{i\pi\omega^E_{i}} \right)
\\
&= 2\pi i \,e^{-2 S_{D}}  \cN_{D}^2 \left( g_s \cdtwo \over \sqrt{2\pi} \right)^{k+1} i^{k+1} 2^{k-1} \cos(2\pi\omega^E)\prod_{i=1}^k\sin(2\pi\omega^E_i) \vphantom{\frac{1}{1}}.
\fe
Finally, analytically continuing back to real energies, we find the $1_R\to k_R$ string amplitude 
\ie
\mathfrak{A}^{D\Dbar}_{1_{R}\to k_{R}}(\omega_1,\ldots,\omega_k)=-e^{-2 S_{D}}  \cN_{D}^2 \left( g_s \cdtwo \over \sqrt{2\pi} \right)^{k+1} 2^{k}\pi \cosh(2\pi\omega)\prod_{i=1}^k\sinh(2\pi\omega_i) \vphantom{\frac{1}{1}}.
\label{eq:1RkRAws}
\fe
This result agrees with the anticipated MQM amplitude (\ref{eq:mmRR1tokleading}) via the dictionary (\ref{mmwszerot}), provided
\ie\label{eq:CD2}
\cdtwo = \frac{\sqrt{2\pi}}{g_s},~~~~~~
\cN_{D} = {1\over 2\pi}.
\fe

The above computation is straightforwardly generalized to the finite temperature case, where we compactify $X=iX^0$ on the thermal circle. The measure factor on D-$\overline{\rm D}$ instanton moduli space at finite temperature is computed in \eqref{eq:DDbarfinT}.  The disc one-point amplitudes \eqref{eq:tachyon1pt} and \eqref{eq:axion1pt} are unmodified, except that the closed string energies are now taken to be Matsubara frequencies, namely $\omega_j^E = {n_j/R}$, for $n_j\in \mathbb{Z}$. 
Altogether, the finite temperature version of (\ref{eq:DDbar1RtokRstep2}) is given by
\ie{}
\mathfrak{A}^{(T), D\Dbar}_{1_{R}\to k_{R}}(\omega_1,\ldots,\omega_k) 
&= 2\pi i \, e^{-2 S_{D}} \cN_{D}^2 \left( g_s \cdtwo \over \sqrt{2\pi} \right)^{k+1} \int_{\widetilde\cC} \frac{d\Delta x}{4R^2\sin\left( \frac{\Delta x -2\pi}{2R} \right)\sin\left( \frac{\Delta x + 2\pi}{2R} \right)} \left( e^{i\pi\omega^E} - e^{-i\omega^E \Delta x}e^{-i\pi\omega^E} \right) 
\\
&~~~~\times \prod_{i=1}^k \left( -e^{-i\pi\omega^E_{i}} + e^{i\omega^E_{i} \Delta x}e^{i\pi\omega^E_{i}} \right) \\
& = 2\pi i \,  e^{-2 S_{D}} \cN_{D}^2 \left( g_s \cdtwo \over \sqrt{2\pi} \right)^{k+1} \frac{i^{k+1} 2^{k} \pi}{R\sin\left(\frac{2\pi}{R}\right)} \cos(2\pi\omega^E)\prod_{i=1}^k\sin(2\pi\omega^E_i),
\fe
where the contour $\widetilde\cC$ now runs around the thermal circle, namely $\Delta x\in [0,2\pi R)$, with a principal value prescription over the singularities of the integrand. This result differs from (\ref{eq:DDbar1RtokRstep2}) simply by the factor ${2\pi\over R \sin {2\pi\over R}}$. This is in precise agreement with the MQM result at finite temperature (\ref{eq:finTfactor}) (recall the convention in conversion, $R_{\rm WS} = 2 R_{\rm MM}$).


\subsection{D-$\overline{\rm D}$ instanton contribution to the uncharged $1_R\to k_L$ amplitude}

The leading order contribution to the uncharged $1_R\to k_L$ scattering process, forbidden in perturbation theory, is also mediated by a D-$\overline{\rm D}$ pair. The corresponding S-matrix is computed by a formulae of the form identical to (\ref{eq:DDbar11}) except that we now insert the vertex operator $\cR^{+}_\omega$ on the first pair of discs, and $\cL^{-}_{\omega_i}$ on the remaining pairs of discs for $i=1,\dots,k$. 
The same steps leading up to \eqref{eq:DDbar1RtokRstep2} now give
\ie\label{eq:DDbar1RtokL}
\mathfrak{A}^{D\overline D}_{1_R\to k_L}(\omega_1,\ldots,\omega_k)
&=2\pi i\, e^{-2 S_D}  \cN_{D}^2 \left( g_s \cdtwo \over \sqrt{2\pi} \right)^{k+1} \int_{\cC} \frac{d\Delta x}{(\Delta x)^2-(2\pi)^2} \left( e^{i\pi\omega^E} - e^{-i\omega^E \Delta x}e^{-i\pi\omega^E} \right)
\\
&~~~~\times \prod_{i=1}^k \left( e^{i\pi\omega^E_{i}} - e^{i\omega^E_{i} \Delta x}e^{-i\pi\omega^E_{i}} \right) \\
& = -2\pi i \, e^{-2 S_D} \cN_{D}^2 \left( g_s \cdtwo \over \sqrt{2\pi} \right)^{k+1} i^{k+1}2^{k-1} \prod_{i=1}^k\sin(2\pi\omega^E_i) \vphantom{\frac{1}{1}}.
\fe
After analytic continuation back to real energies $\omega_j$, taking into account the identification (\ref{eq:CD2}), as well as the dictionary (\ref{mmwszerot}), this is in precise agreement with the anticipated MQM result (\ref{eq:mmTT1tokleading}).




\subsection{Effects of a single charged D-instanton at finite temperature}
\label{chargedfinitews}

Charged processes in the MQM, in which a nonzero net number of fermions (i.e. difference between the numbers of particles versus holes) tunnel from one side of the potential barrier to another, are expected to be dual to string scattering mediated by D-instantons with nonzero net RR charge in type 0B string theory. The analysis of charged processes is complicated by the IR divergence which leads to vanishing exclusive scattering amplitudes. Indeed, as seen in section \ref{sec:finiteT}, the charged D-instanton moduli space measure factor, as computed by the exponentiation of the cylinder diagram (\ref{eq:cylDD}), vanishes in the zero temperature limit $R\to\infty$, in agreement with the MQM expectation.

Nonetheless, there appears to be a well-defined notion of charged amplitude at finite temperature from the worldsheet perspective. Here we analyze a single D-instanton contribution to the $1_R \to k_L + k'_R$ string amplitude at finite temperature, computed by the diagram
\ie\label{smatrixchfint}
e^{-S_D}\exp \Big(\tikz[baseline={([yshift=-1.6ex]current bounding box.center)}]{\pic{cylnnrr0={}{}}} \Big) \int dx_1 ~ \tikz[baseline={([yshift=-1.6ex]current bounding box.center)}]{\pic{disk1r={1}}} \cdot \Big( \tikz[baseline={([yshift=-1.6ex]current bounding box.center)}]{\pic{disk1r={1}}} \cdots  \tikz[baseline={([yshift=-1.6ex]current bounding box.center)}]{\pic{disk1r={1}}} \Big) \cdot \Big(
\tikz[baseline={([yshift=-1.6ex]current bounding box.center)}]{\pic{disk1r={1}}} \cdots \tikz[baseline={([yshift=-1.6ex]current bounding box.center)}]{\pic{disk1r={1}}} \Big),
\fe
where $x_1$ is the Euclidean time coordinate of the D-instanton, and the integrand involves the product of $1+k+k'$ discs. The first disc in the product comes with the insertion of the right asymptotic in-state vertex operator $\cR^{+}_\omega$. The next $k$ pairs of discs come with the insertion of left out-state vertex operators
$\cL^{-}_{\omega_{i,L}}$ for $i=1,\ldots,k$, while the remaining $k'$ pairs of discs come with the insertion of
 $\cR^{-}_{\omega_{j,R}}$ for $j=1,\ldots,k'$.

Using the disc one-point amplitudes \eqref{eq:tachyon1pt}, \eqref{eq:axion1pt}, the cylinder amplitude (\ref{eq:cylDD}), the normalization constants  (\ref{eq:CD2}), and factoring out $(-iR)$ from (\ref{smatrixchfint}) as in the definition of the thermal string amplitude, we obtain 
\ie
{\widetilde{\mathfrak A}}^{(T), D}_{1_R\to k_L + k'_R}(\{\omega_{i,L}\},\{\omega_{j,R}\}) =  e^{-S_D} {i\over R} (-)^{k'} e^{2\pi \omega_L},
\label{eq:WSsingleDamp}
\fe
where $\omega_L \equiv \sum_{i=1}^{k}\omega_{i,L}$. Similarly, a charge $-1$ $\overline{\rm D}$-instanton correction to the $1_R \to k_L + k_R$ scattering amplitude evaluates to
\ie
{\widetilde{\mathfrak A}}^{(T), \overline{D}}_{1_R\to k_L + k'_R}(\{\omega_{i,L}\},\{\omega_{j,R}\}) = e^{-S_D} {i\over R} (-)^{1+k} e^{-2\pi \omega_L}.
\label{eq:WSsingleDbaramp}
\fe
It is tempting to interpret these results as dual to some kind of thermal amplitude for charged processes in the MQM. However, we do not understand the precise definition of such observables from the matrix model perspective.


\section{Spacetime dimensional regularization}
\label{sec:DimReg}

In this section, we describe a dimensional regularization scheme that amounts to deforming the worldsheet matter SCFT to describe strings propagating in $2+\epsilon$ dimensional spacetime (including the Liouville direction). While this scheme may seem unphysical, it allows for a consistent computation of IR safe inclusive scattering probabilities of type 0B string theory in the charged sector (recovered in the $\epsilon\to 0$ limit).


The unitarity relation of the type 0B S-matrix is expressed in its graded subcomponents as 
\ie\label{ssunitch}
\sum_{\ell\in\mathbb{Z}} \cS_{\ell} \cS_{\ell}^\dagger = \mathds{1}.
\fe
For instance, taking the $1\to 1$ matrix element of (\ref{ssunitch}) at order $e^{-2S_D}$ implies that 
\ie
2{\mathfrak A}^{D\overline{D},(0)}_{1\to 1}(\omega) + \SumInt_{\alpha} 
\left( \left| {\mathfrak A}^{D,(0)}_{1\to \alpha}(\omega_\alpha) \right|^2 + \left| {\mathfrak A}^{\overline{D},(0)}_{1\to\alpha}(\omega_\alpha) \right|^2\right) \delta\left( \omega - \omega_\alpha \right) = 0,
\label{eq:1to1unitarity}
\fe
where in the second term we sum over all possible out asymptotic multi-particle states $\ket{\alpha}$ with total energy $\omega_\alpha$, and integrate over their frequency phase space. 

The IR divergence is reflected in the fact that the single D-instanton amplitude (\ref{eq:WSsingleDamp}) and (\ref{eq:WSsingleDbaramp}) have a vanishing overall coefficient in the zero temperature limit, while the energy dependence of the remaining part of the amplitude would lead to a divergent phase space integral with the measure ${d\omega_i\over \omega_i}$ for each out-particle of energy $\omega_i$,\footnote{The phase space measure factor ${1\over \omega_i}$ is due to our normalization convention for 1-particle states, $\langle\omega|\omega'\rangle=\omega\delta(\omega-\omega')$.} and thus naively one runs into a $0\times \infty$ ambiguity in the charged sector contribution to (\ref{eq:1to1unitarity}).\footnote{At finite temperature, one could formally regularize this frequency space integral by replacing it with a sum over imaginary Matsubara frequencies $\omega_j=\frac{i n_j}{R}$ with $n_j\in\bZ_{>0}$, but it is not evident how such a calculation relates to physical scattering probabilities.} In the dimensional regularization scheme, this IR divergence will be tamed in a way that manifestly preserves unitarity. 

\begin{table}
\centering
\begin{tabular}{|c c c c|}
\hline
&&&\\
& \underline{$X^{0}, \psi^{0}$} & $~~~$& \underline{$\cN=(1,1)$ Liouville} \\
&&&\\
$~~~~c~~=$ & $3\over 2$ & & $\frac{3}{2}\left(1+2Q^2\right) = \frac{27}{2}$ \\
&&&\\
& $\downarrow$ & & $\downarrow$\\
&&&\\
& $\frac{3}{2}d$ & & $\frac{3}{2}\left(1+2\left(\frac{9-d}{2}\right)^2\right) = \frac{3}{2}(10-d)~~~~$ \\
&&&\\
\hline
\end{tabular}
\caption{Deformation of central charges in the matter sectors of the worldsheet SCFT in the dimensional regularization scheme. 
}
\label{table:dimreg}
\end{table}

To proceed, we deform the matter sector of the worldsheet SCFT as follows.  The time-like free boson and its Majorana fermion superpartner $(X^0,\psi^0)$, whose central charge is $c=\tilde{c}=\frac{3}{2}$, are replaced formally by a system of $d=1+\epsilon$ free bosons together with their fermion superpartners, with a total central charge $c=\tilde{c}=\frac{3}{2}d$.  At the same time, the background charge $Q$ of the $\cN=1$ Liouville theory is adjusted to $Q=\frac{9-d}{2}$, so that the total central charge of the matter SCFT remains $c=\widetilde c=15$ and the worldsheet theory is BRST invariant or free of Weyl anomaly. This is summarized in Table \ref{table:dimreg}.

A basic consequence of this modification of the worldsheet theory is that the mass squared of the (NS, NS) tachyon is shifted to $m_T^2 = \frac{1-d}{8}$, whereas the (R, R) axion remains massless $m_A^2=0$, as can be seen from the on-shell condition for their respective vertex operators. The worldsheet computations of scattering amplitudes, both in closed string perturbation and D-instanton perturbation theory, are straightforwardly extended to the deformed background. The D-instantons are defined by Dirichlet boundary condition with respect to all $d$ target spacetime dimensions, and the ZZ boundary condition in the Liouville sector. The space of asymptotic closed string states is modified due to both the propagation and polarization in the extra $d-1=\epsilon$ dimensions. Physical observables of type 0B string theory will be recovered in the $\epsilon\to 0$ limit.

Details of D-instanton calculus analogous to that of sections \ref{Dinst} and \ref{sec:dinstuncharged} in dimensional regularization are presented in Appendix \ref{sec:DimRegApp}. Here we summarize some key results. The exponentiated cylinder diagram between a D- and a $\overline{\rm D}$-instanton evaluates to
\ie
\exp  \Big(2 ~\tikz[baseline={([yshift=-1.6ex]current bounding box.center)}]{\pic{cylnnrb0={1}{2}}}  \Big)
=  {\cN_{D\overline{D}}^{\text{d.r.}} \over \Delta x^2 - (2\pi)^2},
\label{eq:measure_dimreg}
\fe
where the normalization constant $\cN_{D\overline{D}}^{\text{d.r.}}$ is given by
\ie
\cN_{D\overline{D}}^{\text{d.r.}} = e^{\frac{2}{\epsilon} - \gamma_E} \, 8\pi.
\label{eq:NDDbar_dimreg}
\fe
The cylinder diagram with both boundary components on the same D-instanton (c.f. (\ref{eq:cylDD})) has no moduli dependence, and gives a finite constant $\cN_{D}^{\text{d.r.}}$ after regularizing the open string divergence. As before, this constant cannot be fixed a priori in the on-shell worldsheet formalism. 
Consistency with (\ref{eq:measure}) requires $\lim_{\epsilon\to 0} ({\cal N}_D^{\rm d.r.})^2{\cal N}_{D\overline{D}}^{\rm d.r.} = {\cal N}_D^2$.

For the purpose of analyzing the unitarity relation (\ref{eq:1to1unitarity}), it is more convenient to work with the tachyon-axion basis of asymptotic states (\ref{eq:vops}) rather than the ${\cal L}$/${\cal R}$ basis (\ref{eq:LRvertexop}), as the IR divergence in the $\epsilon\to 0$ limit will come entirely from the RR sector.\footnote{Our normalization convention here is such that the asymptotic tachyon and axion 1-particle states correspond to the vertex operators in (\ref{eq:vops}) rescaled by $\sqrt{2}$.}
The first term on the LHS of (\ref{eq:1to1unitarity}) takes the form
\ie
& {\mathfrak A}^{D\overline{D},(0)}_{1_{T/A} \to 1_{T/A}}(k) = -e^{-2S_D} (\cN_D^{\text{d.r.}})^2  \cN_{D\overline{D}}^{\text{d.r.}} \, f_{T/A}^{(d)}(k), 
~~~~~~ {\mathfrak A}^{D\overline{D}}_{1_{T/A} \to 1_{A/T}} =0,
\label{eq:1to1TAbasis}
\fe
where the asymptotic particle is labeled by its $d$-dimensional energy-momentum vector $k^{\mu}=(k^0,\vec{k})$. The functions $f_T^{(d)}(k)$ and $f_A^{(d)}(k)$, obtained from integrating the disc diagrams over the D-$\overline{\rm D}$ instanton moduli space, reduces in the $d\to 1$ limit to
\ie
\lim_{d\to 1} f_T^{(d)}(k) = 4\pi \sinh^2(\pi k^0) \sinh(2\pi k^0),~~~~~~ \lim_{d\to 1} f_A^{(d)}(k) = 4\pi \cosh^2(\pi k^0) \sinh(2\pi k^0).
\fe
The sum/integration over out-states in (\ref{eq:1to1unitarity}) involve the single D- or $\overline{\rm D}$-instanton amplitudes, such as
\ie
{\mathfrak A}^{D,(0)}_{1_A\to n_T + n'_A}(\{k_i\},\{k'_j\}) &= (-1)^{n'}e^{-S_{D}}\cN_{D}^{\text{d.r.}} i (2\pi)^d
 \frac{1}{\sqrt{\pi}}\Psi^{\rsect,d}(P) \prod^{n}_{i=1}\frac{1}{\sqrt{\pi}}\Psi^{\nssect,d}(P_i) \prod^{n'}_{j=1}\frac{1}{\sqrt{\pi}}\Psi^{\rsect,d}(P'_j) .
\label{eq:1tonAnT}
\fe
Here, the Liouville momenta $P_i$ are related to $k_i^\mu$ by the mass-shell condition $P_i^2 = (k_i^0)^2 - \vec{k}_i^2 -m_T^2 $, and similarly $P_j'^2 = (k_j'^0)^2 - \vec{k}_j'^2 -m_A^2 $. The Liouville disc one-point functions $\Psi^{\nssect/\rsect,d}$ are given in (\ref{eq:1ptsdimreg}). 
With these results, one can verify that unitarity relations such as
\ie
0 &= 2\, {\mathfrak A}^{D\overline{D},(0)}_{1_A\to 1_A}(k) \vphantom{\sum} \\
& ~~~ +\sum_{n=0}^{\infty}\frac{1}{n!}\sum_{n'=0}^{\infty}\frac{1}{n'!} \prod_{i=1}^{n} \int_0^\infty dk_i^0 \int_{-\infty}^\infty d^{d-1}\vec{k}_i \int_0^\infty dP_i \, 2 \delta\left( (k^0_i)^2-(\vec{k}_i^2+P_i^2+m_T^2) \right) 
\\
& ~~~~~ \times \prod_{j=1}^{n'} \int_0^\infty dk_j'^0 \int_{-\infty}^\infty d^{d-1}\vec{k}'_j \int_0^\infty dP'_j \,2\delta\left( (k'^0_j)^2-(\vec{k}_j'^2+P_j'^2+m_A^2) \right)  \\
& ~~~~~ \times \left( \left|{\mathfrak A}^{D,(0)}_{1_A\to n_T + n'_A}(\{k_i\},\{k'_j\})\right|^2 + \left|{\mathfrak A}^{\overline{D},(0)}_{1_A\to n_T + n'_A}(\{k_i\},\{k'_j\})\right|^2 \right) \delta^{(d)} \left(k^\mu - \sum^{n}_{i=1}k_i^\mu -\sum^{n'}_{j=1}k_j'^\mu \right),
\label{eq:unitaritydimreg}
\fe
are indeed satisfied (see Appendix \ref{sec:DimRegApp} for the details). 

\section{Discussion}
\label{sec:discuss}

In this paper, we have studied the effects of D-instantons in 2D type 0B string theory constructed from ZZ boundary condition of $(1,1)$ type in ${\cal N}=1$ Liouville theory, and found results in perfect agreement with the expected matrix model dual for the uncharged $1\to k$ processes at order $e^{-2\pi\mu}$. The IR divergence of the charged processes can be understood, from the worldsheet perspective, through the cylinder diagram that corrects the measure on the moduli space of a single charged D-instanton. In contrast, the measure on the uncharged D-$\overline{\rm D}$ instanton moduli space is free of such divergences.

Due to the vanishing effective coupling in the asymptotic region, there is a well-defined notion of thermal S-matrix in type 0B string theory. We defined the thermal amplitude $\widetilde{\cal A}^{(T)}$ at Matsubara frequencies in the MQM by an LSZ-like limiting procedure, in which we start from a connected reduced thermal S-matrix element ${\cal A}^{(T)}$, analytically continue to Matsubara frequencies, and strip off the divergent Bose-Einstein statistical factors. The result for the uncharged thermal amplitude is found to be in agreement with the worldsheet computation of string amplitude $\mathfrak{A}^{(T)}$ at finite temperature, where the worldsheet CFT includes the compact boson on the thermal circle, and the vertex operators involve momentum modes on the thermal circle corresponding to the Matsubara frequencies.

There also appear to be well-defined worldsheet computations of charged amplitudes at finite temperature. While these are expected to be dual to thermal observables in the MQM that capture charged transition amplitudes, the precise definition of the latter has thus far eluded us.

There are two ingredients in our D-instanton calculus that are not determined from first principle. One is the finite part of the counter term that enters the regularization of the logarithmic divergence of the cylinder diagram, which we assumed to be a constant that is fixed by comparison with the dual MQM. The other ingredient is the principal value contour prescription in the integration over the D-$\overline{\rm D}$ instanton moduli space. It is likely that both can be derived from open+closed string field theory along the same lines as the analysis of \cite{Sen:2021qdk}.

There are two obvious possible extensions of this work in the near future: subleading orders in $\mu^{-1}$, and subleading orders in $e^{-\pi\mu}$. A direct worldsheet computation of the former, say at order $e^{-2\pi\mu}\mu^{-1}$ for the uncharged process, requires ${\cal N}=1$ super-Virasoro conformal blocks for sphere 4-point function and torus 2-point function. While the recursion formula for the ${\cal N}=1$ sphere 4-point blocks have been numerically implemented recently in \cite{Balthazar:2022atu}, its torus analog is yet to be worked out. As for the extension to higher instanton numbers, it would presumably be necessary to take into account D-instantons constructed from the higher $(m,n)$ ZZ boundary conditions of ${\cal N}=1$ Liouville theory, analogously to \cite{Balthazar:2019ypi}.

Finally, let us remark that both the notion of thermal S-matrix defined in this paper, as well as the mechanism for IR divergence in type 0B string theory/MQM, may find interesting applications to other quantum systems.

\section*{Acknowledgements}

We would like to thank Panos Betzios for discussions. 
XY thanks Cargese Summer Institute, Aspen Center for Physics, Massachusetts Institute of Technology, VR and XY thank Kavli Institute for Theoretical Physics, for their hospitality during the course of this work.
This work is supported in part by a Simons Investigator Award from the Simons Foundation, by the Simons Collaboration Grant on the Non-Perturbative Bootstrap, and by DOE grants DE-SC0007870 and DE-SC0009924.

\appendix

\section{Some details on dimensional regularization}
\label{sec:DimRegApp}

In this appendix we provide some details on the computation of the cylinder diagram between a pair of D- and $\overline{\rm D}$-instantons in dimensional regularization, and the verification of the unitarity relation for the $1_A\to 1_A$ scattering amplitude (\ref{eq:unitaritydimreg}).

\bigskip

{\noindent\bf D-$\overline{\rm D}$ annulus diagram }

The annulus diagram between a D-instanton at Euclidean time $x_1$ and a $\overline{\rm D}$-instanton at $x_2$, which corrects the integration measure on D-instanton moduli space, takes the following form in dimensional regularization (with $q=e^{-2\pi t}$),
\ie
\tikz[baseline={([yshift=-1.6ex]current bounding box.center)}]{\pic{cylnnrb0={1}{2}}}  
~=& \int_0^\infty \frac{dt}{2t} \frac{1}{2}\left[ \left( \frac{\theta_3(it)}{\eta(it)^3} \right)^{1\over 2} q^{\frac{d-1}{16}}\left( q^{- \frac{1}{2}} - 1 \right) \left( \frac{\theta_3(it)}{\eta(it)} \right)^{d\over 2} \frac{e^{-\left(\frac{\Delta x}{2\pi}\right)^2 \pi t}}{\eta(it)^d} \frac{\eta(it)}{\theta_3(it)} \eta(it)^2\right. \\
& ~~~~~~~~~~ +\left. \left( \frac{\theta_4(it)}{\eta(it)^3} \right)^{1\over 2} q^{\frac{d-1}{16}}\left( q^{- \frac{1}{2}} + 1 \right) \left( \frac{\theta_4(it)}{\eta(it)} \right)^{d\over 2} \frac{e^{-\left(\frac{\Delta x}{2\pi}\right)^2 \pi t}}{\eta(it)^d} \frac{\eta(it)}{\theta_4(it)} \eta(it)^2\right] \\
= &\int_0^\infty \frac{du}{2} u^{-\frac{d}{2}} e^{-\left(\frac{\Delta x}{2\pi}\right)^2 \frac{\pi}{u}} \left[ \int_0^\infty \frac{dP}{\pi} \left( \Psi^{\nssect,d}(P) \right)^{2} e^{-\pi u P^2} \left( \frac{\theta_3(iu)}{\eta(iu)^3} \right)^{\frac{d-1}{2}} \right. \\
& ~~~~~~~~~~~~~~~~~~~~~~~~~~ +\left. \int_0^\infty \frac{dP}{\pi} \left( \Psi^{\rsect,d}(P) \right)^{2} e^{-\pi u P^2} 2^{\frac{1-d}{2}}\left( \frac{\theta_2(iu)}{\eta(iu)^3} \right)^{\frac{d-1}{2}} \right].
\label{eq:cylDDbar_dimreg}
\fe
The second equality expresses the result as an expansion in the closed string channel by changing the integration variable $t=1/u$ and modular transforming the integrand. Here
\ie
\left( \Psi^{\nssect,d}(P) \right)^{2} & = 2\pi \frac{1}{2}\left( \cosh\left[ \sqrt{\frac{9-d}{2}} \pi P\right] - \cosh\left[ \sqrt{\frac{1-d}{2}} \pi P\right] \right), \\
\left( \Psi^{\rsect,d}(P) \right)^{2} & = 2\pi \frac{1}{2} \,2^{\frac{d-1}{2}}\left( \cosh\left[ \sqrt{\frac{9-d}{2}} \pi P\right] + \cosh\left[ \sqrt{\frac{1-d}{2}} \pi P\right] \right)
\label{eq:1ptsdimreg}
\fe
are the squares of the $\cN=1$ Liouville disc NSNS and RR one-point functions (\ref{eq:disk1pt11}) with background charge $Q=\frac{9-d}{2}$, subject to the ZZ boundary condition. 
Note that 
only the scalar component out of the $2^{d-1}$ RR sector ground states propagates along the cylinder. 

The divergence at $d=1$ comes from the RR sector, as can be isolated by considering the leading term in the expansion of $\theta_2/\eta^3$ in the RR contribution to the annulus diagram,
\ie
\int_0^\infty & \frac{dP}{\pi} \left( \Psi^{\rsect,d}(P) \right)^{2} \int_0^\infty du \frac{1}{2} u^{-\frac{d}{2}} e^{-\left(\frac{\Delta x}{2\pi}\right)^2 \frac{\pi}{u}} e^{-\pi u P^2} \\
& = \int_0^\infty \frac{dP}{\pi} \left( \Psi^{\rsect,d}(P) \right)^{2} \left(
\left(\frac{\Delta x}{2\pi}\right)^2\frac{1}{P^2} \right)^{\frac{2-d}{4}} K_{\frac{1}{2}(d-2)}\left(|\Delta x|P\right) \\
& = 2^{d-3}(2\pi)^{\frac{d-1}{2}}\left[ \left(\Delta x^2 - \frac{9-d}{2}\pi^2\right)^{\frac{1-d}{2}} + \left(\Delta x^2 - \frac{1-d}{2}\pi^2\right)^{\frac{1-d}{2}} \right] \Gamma\left({d-1\over 2}\right)
\\
& = \frac{1}{\epsilon} +\frac{1}{2}\left( -\gamma_E + \log(8\pi) \right) - \frac{1}{4}\left( \log\left(\Delta x^2\right) + \log\left(\Delta x^2 - (2\pi)^2\right) \right) + O(\epsilon)
\fe
where in the last line we have set $d=1+\epsilon$ and expanded in small $\epsilon$.

Having isolated the divergent piece, we can subtract it from the full RR contribution in (\ref{eq:cylDDbar_dimreg}) and evaluate the rest at $d=1$. This give zero for the RR piece, while for the NSNS piece we can readily compute
\ie
\int_0^\infty \frac{dt}{2t} \frac{1}{2} e^{-\left(\frac{\Delta x}{2\pi}\right)^2 \pi t} \left( e^{\pi t} - 1 \right) = \frac{1}{4} \log\left( \frac{\Delta x^2}{\Delta x^2 - (2\pi)^2} \right).
\fe
Adding the NSNS and RR contributions, we obtain that the exponentiated cylinder diagram between a D-instanton and a $\overline{\rm D}$-instanton in dimensional regularization evaluates to (\ref{eq:measure_dimreg}) with (\ref{eq:NDDbar_dimreg}) as asserted.

\bigskip

{\noindent\bf Verification of $(\ref{eq:unitaritydimreg})$}

The first term in the unitarity relation (\ref{eq:unitaritydimreg}) is twice the D-$\overline{\rm D}$ correction to the $1_A\to 1_A$ scattering amplitude computed in (\ref{eq:1to1TAbasis}), namely 
\ie
- e^{-2S_{D}} (\cN_D^{\text{d.r.}})^2\cN_{D\overline{D}}^{\text{d.r.}}\, 8\pi \cosh(\pi\omega)^2\sinh(2\pi\omega).
\label{eq:term1ddbar}
\fe
The second term  evaluates to
\ie
e&^{-2S_{D}}(\cN_D^{\text{d.r.}})^2 (2\pi)^{2d} \frac{2}{\pi}\Psi^{\rsect,d}(P)^2  \sum_{n=0}^{\infty}\frac{1}{n!}\sum_{n'=0}^{\infty}\frac{1}{n'!} 
\\
&~~~~\times \prod_{i=1}^{n} \int_0^\infty dk_i^0 \int_{-\infty}^\infty d^{d-1}\vec{k}_i \int_{-\infty}^\infty dP_i \,\delta\left( (k^0_i)^2-(\vec{k}_i^2+P_i^2+m_T^2) \right)  \frac{1}{\pi}\Psi^{\nssect,d}(P_i)^2 
\\
&~~~~\times \prod_{j=1}^{n'} \int_0^\infty dk_j'^0 \int_{-\infty}^\infty d^{d-1}\vec{k}_j' \int_{-\infty}^\infty dP'_j \,\delta\left( (k'^0_j)^2-(\vec{k}_j'^2+P_j'^2+m_A^2) \right) \frac{1}{\pi}\Psi^{\rsect,d}(P_j')^2 
\\
&~~~~\times  \delta^{(d)}\left(k^\mu - \sum^{n}_{i=1}k_i^\mu -\sum^{n'}_{j=1}k_j'^\mu \right) .
\label{eq:term2dsum}
\fe
To proceed, it is convenient to Fourier/Laplace transform with respect to $k^\mu$ by integrating the sum in (\ref{eq:term2dsum}) against $\int d^dk\, e^{ik\cdot x}$ for imaginary $x^0$ with sufficiently large $-{\rm Im} (x^0)$. The transformed sum then exponentiates to
\ie
\exp \left[  \int_{-\infty}^{\infty} d^d\vec{p} \, e^{-iE_{\vec{p}} x^0} \frac{1}{E_{\vec{p}}} e^{i\vec{k}\cdot \vec{x}} \frac{\Psi^{\nssect,d}(P)^2}{2\pi} +  \int_{-\infty}^{\infty} d^d\vec{p} \, e^{-iE'_{\vec{p}} x^0} \frac{1}{E'_{\vec{p}}} e^{i\vec{k}\cdot \vec{x}} \frac{\Psi^{\rsect,d}(P)^2}{2\pi} \right]
\label{eq:expNSexpR}
\fe
where we have performed the $k_i^0, k_j'^0$ integration, and defined $\vec{p} = (k^1,\ldots, k^{d-1},P)$, $E_{\vec{p}} = \sqrt{\vec{p}^2+m_T^2}$, $E'_{\vec{p}} = \sqrt{\vec{p}^2+m_A^2}$.

The NS piece of (\ref{eq:expNSexpR}) is not divergent and can be evaluated directly at $d=1$; its exponent evaluates to
\ie
\frac{1}{2} \log\left( -(x^0)^2 \right) - \frac{1}{2}\log\left( -(x^0)^2 - (2\pi)^2 \right).
\label{eq:NSexponent}
\fe
On the other hand, the R sector piece of (\ref{eq:expNSexpR}) diverges at $d=1$. Using (\ref{eq:1ptsdimreg}),
we can write the second term in the exponent of (\ref{eq:expNSexpR}) as the sum of
\ie
2^{\frac{d-1}{2}}\frac{1}{4} 
 \int_{-\infty}^{\infty} d^d\vec{p} \, e^{i\vec{p}\cdot \vec{y}} e^{-i |\vec{p}| x^0} \frac{1}{|\vec{p}|}, ~~~ \text{with } \vec{y} = \left(x^1, \ldots, x^{d-1},y^{(d)}\right),
\label{eq:FTradial}
\fe
where the last component of the newly introduced vector $\vec{y}$ takes the four values
\ie
y^{(d)} = \left\lbrace \pm \pi\sqrt{\frac{9-d}{2}}, \pm \pi\sqrt{\frac{1-d}{2}} \right\rbrace .
\label{eq:yds}
\fe
The integral in (\ref{eq:FTradial}) is just a Fourier transform of a radially symmetric function, and hence it is given by a special case of a Hankel transform,
\ie
2^{\frac{d-1}{2}}\frac{1}{4} 
|y|^{\frac{d}{2}-1} (2\pi)^{\frac{d}{2}} \int_0^{\infty} dr\, r J_{\frac{d}{2}-1}(r|y|) r^{\frac{d}{2}-1} \frac{1}{r} e^{-irx^0}.
\fe
Performing the integral and summing over the four cases (\ref{eq:yds}) we obtain
\ie
2^{\frac{3d-5}{2}} \pi^{\frac{d-1}{2}} \Gamma\left( \frac{d-1}{2} \right) \left[ \left( \frac{9-d}{2}\pi^2 - (x^0)^2 + \vec{x}^2\right)^{\frac{1-d}{2}} + \left( \frac{1-d}{2}\pi^2 - (x^0)^2 + \vec{x}^2\right)^{\frac{1-d}{2}}  \right].
\fe
We set $d=1+\epsilon$ and expand in small $\epsilon$ to obtain the result
\ie
\frac{2}{\epsilon} - \gamma_E + \log(8\pi) -  \frac{1}{2} \log\left( -(x^0)^2 \right) - \frac{1}{2}\log\left( -(x^0)^2 - (2\pi)^2 \right).
\label{eq:Rexponent}
\fe
Combining (\ref{eq:NSexponent}) and (\ref{eq:Rexponent}), we obtain that (\ref{eq:expNSexpR}) evaluates to
\ie
e^{\frac{2}{\epsilon}-\gamma_E} 8\pi\frac{1}{-(x^0)^2-(2\pi)^2}.
\fe
Laplace transforming back using 
\ie
\frac{1}{-(x^0)^2-(2\pi)^2} \to \frac{1}{2\pi} \sinh(2\pi\omega),
\fe
we finally obtain that the second contribution is equal to 
\ie
e^{-2S_D}(\cN_D^{\text{d.r.}})^2 e^{\frac{2}{\epsilon}-\gamma_E} (2\pi)^2 2^4 \cosh(\pi\omega)^2 \sinh(2\pi\omega),
\fe
which precisely cancels the first contribution from a D-$\overline{\rm D}$ pair (\ref{eq:term1ddbar}), 
restoring unitarity at order $e^{-2S_D}$.

\bibliographystyle{JHEP}
\bibliography{0B_refs}

\end{document}